\documentclass[a4paper,11pt]{article}
\usepackage{heppub}
\pdfoutput=1
\usepackage{amsmath} 
\usepackage{amssymb} 
\usepackage{graphicx}
\usepackage{hyperref}
\usepackage{pgfplots} 
\pgfplotsset{compat=1.18}
\usepackage{slashed} 
\usepackage{soul} 
\usepackage{xcolor} 

\newcommand{\DS}{\displaystyle} 
\newcommand{\EI}{} 
\renewcommand{\L}{\left}
\newcommand{\R}{\right}
\newcommand{\<}{\!} 
\renewcommand{\>}{\,} 

\newcommand{\Biggwhite}{{\color{white} \Bigg|}}

\newcommand{\sss}{\mathchoice %
{\displaystyle} %
{\displaystyle} %
{\scriptscriptstyle} %
{\scriptscriptstyle} %
}

\renewcommand{\b}{\bar} 
\newcommand{\h}{\hat} 

\newcommand{\alphaEW}{\alpha_{\sss\rm EW}}
\newcommand{\alphaS}{\alpha_{s}}

\newcommand{\thetaW}{\theta_{W}}

\newcommand{\muV}{\mu_{V}}
\newcommand{\muW}{\mu_{W}}
\newcommand{\muZ}{\mu_{Z}}

\renewcommand{\d}{{\rm d}}

\newcommand{\LO}{{\sss\rm LO}}
\newcommand{\NLO}{{\sss\rm NLO}}
\newcommand{\NNLO}{{\sss\rm NNLO}}

\newcommand{\NLLprime}{{\sss\rm NLL'}}

\newcommand{\NNLLprime}{{\sss\rm NNLL'}}

\newcommand{\Geneva}{{\small\textsc{Geneva}}}

\newcommand{\LHAPDF}{{\small\textsc{LHAPDF}}}

\newcommand{\mb}{m_b}
\newcommand{\mt}{m_t}
\newcommand{\mV}{M_V}
\newcommand{\mW}{M_W}
\newcommand{\mZ}{M_Z}
\newcommand{\mH}{M_H}

\newcommand{\GammaV}{\Gamma_V}
\newcommand{\GammaW}{\Gamma_W}
\newcommand{\GammaZ}{\Gamma_Z}
\newcommand{\GammaH}{\Gamma_H}

\newcommand{\CF}{C_F}
\newcommand{\CA}{C_A}

\newcommand{\nf}{n_f}

\newcommand{\GeV}{{\rm GeV}}

\newcommand{\CS}{{\sss\rm CS}} 
\newcommand{\cut}{{\sss\rm cut}} 
\newcommand{\MAX}{{\sss\rm max}} 
\newcommand{\MC}{{\sss\rm MC}} 
\newcommand{\MIN}{{\sss\rm min}} 
\newcommand{\nonproj}{{\sss\rm nonproj}} 
\newcommand{\NS}{{\sss\rm NS}} 
\newcommand{\OS}{{\sss\rm OS}} 

\graphicspath{{plots/}}

\usepackage{xcolor}
\usepackage{marginnote}
\usepackage{tikz-feynman}
\usepackage{caption}
\usepackage{subcaption}
\definecolor{darkgreen}{rgb}{0.0, 0.4, 0.0}
\usepackage{hyperref}
\hypersetup{
    linktocpage,
     colorlinks,
     citecolor=darkgreen,
     linkcolor= darkgreen,
     urlcolor=darkgreen
}

\newcommand{\ord}[1]{\mathcal{O}(#1)}

\newcommand{\df}{\mathrm{d}}

\newcommand{\nn}{\nonumber}

\newcommand{\one}{{(1)}}

\newcommand{\ptv}{p_{T}^\cut}
\newcommand{\ptva}{p_{1,T}^{\mathrm{cut}}}
\newcommand{\ptvb}{p_{2,T}^{\mathrm{cut}}}
\newcommand{\ptja}{p_{1,T}^{J}}
\newcommand{\ptjb}{p_{2,T}^{J}}
\newcommand{\Tau}{r}

\newcommand{\cP}{\mathcal{P}}

\newcommand{\as}{\alpha_s}

\newcommand{\lqcd}{\Lambda_\mathrm{QCD}}

\newcommand{\dsigMC}{\df\sigma^\textsc{mc}}

\newcommand{\SC}{\mathcal{S}^R}

\newcommand{\SCNG}{\mathcal{S}^\text{NG}}

\newcommand{\Rj}{R_J}

\newcommand{\geneva}{\textsc{Geneva}~}
\newcommand{\scetlib}{{\tt SCETlib}~}
\newcommand{\Matrix}{\textsc{Matrix}~}
\newcommand{\pythia}{\textsc{Pythia8}~}
\newcommand{\openloops}{{\tt OpenLoops}~}

\newcommand{\rescaletwoplots}{0.475\textwidth}

\newcommand{\vspacebetweentwoplots}{2ex}

\newcommand{\spaceabovefigurecaption}{\vspace*{-2ex}}
\newcommand{\spacebelowfigurecaption}{\vspace*{0ex}}
\allowdisplaybreaks[2]

\title{\boldmath  NNLO+PS $W^+W^-$ production using jet veto resummation at NNLL$'$}

\author[a]{Alessandro Gavardi,}
\author[b]{Matthew A. Lim,}
\author[c]{Simone Alioli}
\author[a]{and Frank Tackmann}

\affiliation[a]{Deutsches Elektronen-Synchrotron DESY, Notkestr. 85, 22607 Hamburg, Germany\vspace{0.5ex}}
\affiliation[b]{Department of Physics and Astronomy, University of Sussex, Sussex House, Brighton, BN1 9RH, UK\vspace{0.5ex}}
\affiliation[c]{Universit\`{a} degli Studi di Milano-Bicocca \& INFN, Piazza della Scienza 3, Milano 20126, Italy\vspace{0.5ex}}

\emailAdd{alessandro.gavardi@desy.de}
\emailAdd{m.a.lim@sussex.ac.uk}

\abstract{%
We construct a novel event generator for the process \mbox{$p \> p \to \ell^- \> \b{\nu}_\ell \> \ell'^+ \> \nu_{\ell'}$}, which matches fixed-order predictions at next-to-next-to-leading order in the strong coupling to a parton shower program. The matching is achieved using the \textsc{Geneva} method, in this case exploiting a resummed calculation for the hardest jet transverse momentum at next-to-next-to-leading logarithmic accuracy obtained via soft-collinear effective theory and implemented in the \texttt{C++} library \texttt{SCETlib}. This choice of resolution variable ensures that the introduction of a jet veto, commonly used by experimental analyses to reject multi-jet background events, does not result in the appearance of unmitigated large logarithms for low veto scales before showering. After validating our partonic results against publicly available fixed order and resummed calculations, we compare our predictions to measurements taken at the ATLAS and CMS experiments, finding good agreement. This is the first NNLO+PS accurate event generator to use the hardest jet transverse momentum as a resolution variable.
}

\begin{document}

\maketitle

\section{Introduction}
\label{sec:intro}

The availability of precise Monte Carlo event generators is a crucial component for the success of the LHC physics programme at run III and beyond. Though advances in the accuracy of theoretical predictions have been made on many fronts, one particular area which has seen much fruitful progress has been in the matching of next-to-next-to-leading order (NNLO) calculations to parton shower (PS) programs, which goes by the acronym NNLO+PS. Of the three major approaches currently under development~\cite{Alioli:2012fc,Monni:2019whf,Campbell:2021svd}, the \geneva method is unique in exploiting resummed calculations at state-of-the-art accuracy to achieve the matching. For colour-singlet processes, resummation of the $0-$jettiness variable~\cite{Stewart:2009yx,Stewart:2010tn} up to N$^3$LL has been studied in the context of NNLO+PS generators~\cite{Alioli:2015toa,Alioli:2019qzz,Alioli:2020qrd,Alioli:2020fzf,Alioli:2021egp,Cridge:2021hfr,Alioli:2022dkj,Alioli:2023har}, while the resummation formalism for the same variable in top-quark pair production has been studied in ref.~\cite{Alioli:2021ggd}. In addition, \geneva has made use of colour-singlet transverse momentum resummation up to N$^3$LL~\cite{Alioli:2021qbf}, provided by the standalone code \textsc{RadISH}~\cite{Monni:2016ktx,Bizon:2017rah}. 

In this work, we instead consider using the resummation of the hardest jet transverse momentum, $\ptja$, to construct an NNLO+PS accurate event generator. This is an interesting avenue to pursue for several reasons. Firstly, it is common for experimentalists to apply a jet veto in order to suppress backgrounds with jets in the final state. For example, measurements of $H\to W^+W^-$ commonly use a veto on jets to remove contamination from the $t\bar{t}$ process, which has a similar final state. In placing the veto, large logarithms of the ratio of the veto scale $\ptv$ to the `hard', high-energy scale $Q$ arise, and can be sufficiently large as to threaten the convergence of the perturbation theory. This motivates the resummation of these logarithms so that one may obtain reliable predictions for the exclusive zero-jet cross section. Embedding this resummation inside a Monte Carlo event generator would extend the resummed calculation to allow fully exclusive, high multiplicity final states to be simulated. Secondly, the construction of NNLO+PS generators normally relies on the choice of a resolution variable. This can be considered to be a source of theoretical uncertainty, since for exclusive observables the predictions given by generators using different choices of said variable may differ considerably~\cite{Alioli:2021qbf}. It is therefore important to be able to assess the effect of the resolution variable choice on NNLO+PS predictions and to explore several different options. 

To these ends, we have constructed a \geneva implementation utilising
jet veto resummation for the production of a pair of leptonically-decaying opposite-sign $W$ bosons. Previous NNLO+PS implementations of this process have appeared in refs.~\cite{Re:2018vac,Lombardi:2021rvg} using the colour-singlet transverse momentum $q_T$ as a resolution variable: this is, however, a process where the effects of jet veto resummation are extremely important, and hence provides an interesting case study for our new approach. We stress, however, that the formalism which we develop in this work is applicable to all colour-singlet production processes. 

Specifically, the observable we consider is
\begin{align}
\ptja=\max_{m\in J(R)} |\vec{p}_{m,T}|
\end{align}
where the index $m$ runs over all jets in the final state obtained via a clustering algorithm. Defining the second hardest jet $\ptjb$ in a similar way, we then define jet bins via
\begin{align}
\label{eq:NNLOevents}
  \text{$\Phi_0$ events: }
& \qquad \frac{\dsigMC_0}{\df\Phi_0}\left(\ptva\right)
\,,\nn\\
\text{$\Phi_1$ events: }
& \qquad
\frac{\dsigMC_{1}}{\df\Phi_{1}}\left(\ptja > \ptva; \ptvb\right)
\,,
\\
\text{$\Phi_2$ events: }
& \qquad
\frac{\dsigMC_{\ge 2}}{\df\Phi_{2}}\left(\ptja > \ptva, \ptjb > \ptvb\right)
\,. \nn
\end{align}
In principle, the two parameters $\ptva$ and $\ptvb$ which set the veto scales (below which jets are vetoed) can be taken to be different. This approach to defining IR-finite events at NNLO, together with the use of a resummed calculation at NNLL$'$, is the core of the \geneva method, first introduced in ref.~\cite{Alioli:2012fc}. 

The formalism for the resummation of jet veto logarithms in the case of colour-singlet production has been developed using both direct QCD~\cite{Banfi:2012jm,Banfi:2012yh,Banfi:2015pju} and soft-collinear effective theory (SCET)~\cite{Becher:2012qa,Tackmann:2012bt,Becher:2013xia,Stewart:2013faa,Becher:2014aya} methods. A number of different processes have been studied -- in particular, gauge boson pair production has been examined in several works~\cite{Moult:2014pja,Monni:2014zra,Wang:2015mvz,Dawson:2016ysj,Arpino:2019fmo}. In ref.~\cite{Monni:2019yyr}, a double-differential resummation for Higgs boson production was achieved, simultaneously resumming logarithms of the boson transverse momentum and the veto logarithms. In the case of colour singlets produced in association with one or more hard jets, factorisation and resummation at NLL$'$ were achieved in refs.~\cite{Liu:2012sz,Liu:2013hba}.

A number of public tools to perform jet veto resummation in the colour-singlet case now exist. The JetVHeto~\cite{Banfi:2012jm,Banfi:2015pju} and RadISH programs exploit the CAESAR formalism for resummation~\cite{Banfi:2004yd,Banfi:2012yh} and its extension to NNLL~\cite{Banfi:2012jm}. The former has been interfaced to fixed-order predictions in the code MCFM-RE~\cite{Arpino:2019fmo}, while the latter has been interfaced to the NNLO provider \Matrix~\cite{Kallweit:2020gva}. Recently, the MCFM collaboration have implemented the resummation in a SCET framework up to N$^3$LL$_{\rm p}$ and have matched this to fixed order calculations for a number of colour-singlet production processes~\cite{Campbell:2023cha}.

The structure of the paper is as follows. In \sec{theory}, we discuss
the theoretical framework of our calculation. We provide a recap of
the \geneva method in \sec{geneva} and introduce an extension thereof which 
includes the NLL$'$ resummation of $\ptjb$ in \sec{NLLp1},
before discussing the resummed calculations needed to separate the jet
bins in \sec{0jet} and \sec{1jet}. We provide further details about
the implementation in \sec{details}, including how we match the
resummed and fixed order predictions and estimate theoretical
uncertainties -- we also discuss the interface to the parton shower.
In \sec{results}, we first validate our fixed order
and resummed results against the public codes \Matrix and MCFM 
  (\sec{validation} and \sec{MCFM_validation}). We then present our
showered results in \sec{showeredresults}, before comparing to
experimental data from the ATLAS and CMS experiments in
\sec{expdata}. We present our conclusions in \sec{conclusions}, and
offer suggestions for future extensions of this work.
\section{Theoretical framework}
\label{sec:theory}
In this section we briefly recall the \geneva method for defining events at NNLO. We then discuss in more detail the resummed calculation which we utilise in this work, namely the jet veto resummation for colour-singlet production obtained via a SCET approach. We detail our treatment of the matching of fixed order and resummed calculations, as well as the procedure we follow in order to obtain an estimate of the theoretical uncertainties, and then cover the matching of the partonic calculation to the parton shower. 

\subsection{The G{\scriptsize ENEVA} method}
\label{sec:geneva}
In this section, we recall the \geneva formul\ae ~for the Monte Carlo partonic jet bins, which are defined using two generic resolution variables $r_N$ and $r_{N+1}$. The full derivation of these expressions can be found in e.g. refs.~\cite{Alioli:2015toa,Alioli:2019qzz}.

\geneva relies on a partitioning of the partonic event space into three regions: the $\Phi_N$ phase space contains events with no extra emissions (beyond those potentially present at Born level), while the $\Phi_{N+1}$ and $\Phi_{N+2}$ bins contain events with one and two additional partonic jets respectively. The thresholds between regions are denoted by $\Tau_N^\cut$ and $\Tau_{N+1}^\cut$.

For the case of colour singlet production, the differential cross section for the production of events with zero additional emissions is given by
\begin{align}
\frac{\dsigMC_0}{\df\Phi_0}(\Tau_0^\cut) &= \frac{\df\sigma^{\rm NNLL'}}{\df\Phi_0}(\Tau_0^\cut) - \frac{\df\sigma^{\rm NNLL'}}{\df\Phi_{0}}(\Tau_0^\cut)\bigg\vert_{\NNLO_0} \nn \label{eq:0full}\\
&\qquad +(B_0+V_0+W_0)(\Phi_0)\, +\,  \int \frac{\mathrm{d} \Phi_1}{\mathrm{d} \Phi_0} (B_1 + V_1)(\Phi_1)\,\theta\big( \Tau_0(\Phi_1)< \Tau_0^{\mathrm{cut}}\big) \nn \\
&\qquad+  \int \frac{\mathrm{d} \Phi_2}{\mathrm{d} \Phi_0} \,B_2 (\Phi_2) \, \theta\big( \Tau_0(\Phi_2)< \Tau_0^{\mathrm{cut}}\big)\nn \\
&= \frac{\df\sigma^{\rm NNLL'}}{\df\Phi_0}(\Tau_0^\cut) - \frac{\df\sigma^{\rm NNLL'}}{\df\Phi_{0}}(\Tau_0^\cut)\bigg\vert_{\NLO_0} \nn \\
&\qquad +(B_0+V_0)(\Phi_0)\, +\,  \int \frac{\mathrm{d} \Phi_1}{\mathrm{d} \Phi_0} B_1(\Phi_1)\,\theta\big( \Tau_0(\Phi_1)< \Tau_0^{\mathrm{cut}}\big)\nn\\
&\qquad + \mathcal{O}\left(\as^2 \Tau_0^\cut\right)\,,
\end{align}
where in the second equality we have dropped $\mathcal{O}(\as^2)$ power corrections in $\Tau_0^\cut$. For the case of a single extra emission we have two contributions: that above $\Tau_0^\cut$,
\begin{align}
\frac{\dsigMC_{1}}{\df\Phi_{1}} (\Tau_0 > \Tau_0^\cut; \Tau_{1}^\cut) &= \Bigg\{\Bigg[ \frac{\df\sigma^{\rm NNLL'}}{\df\Phi_0\df\Tau_0}-\frac{\df\sigma^{\rm NNLL'}}{\df\Phi_0\df\Tau_0}\bigg\vert_{\NLO_1}\,\Bigg]\, \cP_{0\to 1}(\Phi_1)\,+ (B_1 + V_1^C)(\Phi_1)  \Bigg\} \nn \label{eq:1masterfull}\\
&\qquad\times\, U_1(\Phi_1, \Tau_1^\cut)\, \theta(\Tau_0 > \Tau_0^\cut) \nn \\
&\qquad+\int\ \bigg[\frac{\df\Phi_{2}}{\df\Phi^\Tau_1}\,B_{2}(\Phi_2) \, \theta\!\left(\Tau_0(\Phi_2) > \Tau_0^\cut\right)\,\theta(\Tau_{1} < \Tau_1^\cut) \nn \\
&\qquad\quad - \frac{\df\Phi_2}{\df \Phi^C_1}\, C_{2}(\Phi_{2}) \, \theta(\Tau_0 > \Tau_0^\cut) \bigg] \nn \\
&\qquad- B_1(\Phi_1)\, U_1^\one(\Phi_1, \Tau_1^\cut)\, \theta(\Tau_0 > \Tau_0^\cut)\,,
\end{align}
and the nonsingular piece below $\Tau_0^\cut$, which arises from non-projectable configurations,
\begin{align}
\frac{\dsigMC_{1}}{\df\Phi_{1}} (\Tau_0 \le \Tau_0^\cut; \Tau_{1}^\cut) &= (B_1+V_1)(\Phi_1)\, \overline{\Theta}^{\mathrm{FKS}}_{\mathrm{map}}(\Phi_1) \, \theta(\Tau_0<\Tau^{\mathrm{cut}}_0)\nn \\
&= B_1(\Phi_1)\, \overline{\Theta}^{\mathrm{FKS}}_{\mathrm{map}}(\Phi_1) \, \theta(\Tau_0<\Tau^{\mathrm{cut}}_0)+ \mathcal{O}\left(\as^2 \Tau_0^\cut\right)\,.
 \label{eq:1belowtau0}
\end{align}
In a similar vein the case of two extra emissions also receives two contributions,
\begin{align}
\frac{\dsigMC_{\geq 2}}{\df\Phi_{2}} (\Tau_0 > \Tau_0^\cut, \Tau_{1}>\Tau_{1}^\cut) &= \Bigg\{ \bigg[ \frac{\df\sigma^{\rm NNLL'}}{\df\Phi_0\df\Tau_0} - \frac{\df\sigma^{\rm NNLL'}}{\df\Phi_0\df\Tau_0}\bigg|_{\NLO_1}\bigg]\, \cP_{0\to 1}(\widetilde{\Phi}_1) \nn \\
&\qquad + (B_1 + V_1^C)(\widetilde{\Phi}_1)\Bigg\} \,  U_1'(\widetilde{\Phi}_1, \Tau_1)\, \theta(\Tau_0 > \Tau_0^\cut) \Big\vert_{\widetilde{\Phi}_1 = \Phi_1^\Tau\!(\Phi_2)} \nn\\
&\qquad\quad\times\, \cP_{1\to 2}(\Phi_2) \, \theta(\Tau_1 > \Tau_1^\cut)
\nn \\
&\qquad + \Big\{ B_2(\Phi_2)\, \theta(\Tau_{1}>\Tau^{\mathrm{cut}}_{1})- B_1(\Phi_1^\Tau)\,U_1^{\one\prime}\!\big(\widetilde{\Phi}_1, \Tau_1\big)\nn\\
&\qquad\quad\times\,\cP_{1\to 2}(\Phi_2)\, \Theta(\Tau_1 > \Tau_1^\cut)
\Big\}\, \theta\left(\Tau_0(\Phi_2) > \Tau_0^\cut\right)\,, \label{eq:2masterfull}
\end{align}
and
\begin{align}
\frac{\dsigMC_{\geq 2}}{\df\Phi_{2}} (\Tau_0 > \Tau_0^\cut, \Tau_{1} \le \Tau_{1}^\cut) &= B_2(\Phi_2)\, \overline{\Theta}_{\mathrm{map}}^\Tau(\Phi_2) \, \theta(\Tau_1 < \Tau_1^\cut)\, \theta\left(\Tau_0(\Phi_2) > \Tau_0^\cut\right)\,,  \label{eq:2belowtau1}
\end{align}
above and below $\Tau_1^\cut$ respectively.

We denote by $B_n$, $V_n$ and $W_n$ the $0$-, $1$- and $2$-loop
matrix elements for $n$ QCD partons in the final state (including parton densities). The notation $\mathrm{N}^k\mathrm{LO}_n$ indicates a quantity with $n$ additional partons in the final state
computed at $\mathrm{N}^k\mathrm{LO}$ accuracy.

The resummed and resummed-expanded terms of the calculation are differential in the zero-parton phase space, while the fixed order calculation naturally includes contributions from higher multiplicities. It is therefore often necessary to perform a projection from a higher to a lower multiplicity in order to evaluate the resummed/resummed-expanded terms; we denote such a projected phase space point as $\widetilde{\Phi}_{N}$. In addition, we use the notation
\begin{align}
\label{eq:dPhiRatio}
 \frac{\df \Phi_{M}}{\df \Phi_N^{\cal O}}  = \df \Phi_{M} \, \delta[ \widetilde{\Phi}_N - \Phi^{\cal O}_N(\Phi_M) ] \,\Theta^{\cal O}(\Phi_M)
\end{align}
to indicate an integration over the portion of the lower multiplicity $\Phi_M$ phase space which can be reached from a higher multiplicity $\Phi_N$ point while keeping some observable $\cal O$ fixed.
The $\Theta^{\cal O}(\Phi_M)$ term ensures that only phase space points which are genuinely singular in the observable $\cal O$ are considered.
When generating $1$-body events, for example, the relevant shorthand reads
\begin{equation} \label{eq:Phi1TauProj}
\frac{\df\Phi_2}{\df\Phi_1^\Tau} \equiv \df\Phi_2\,\delta[\widetilde{\Phi}_1 - \Phi^\Tau_1(\Phi_2)]\,\Theta^\Tau(\Phi_2)
\,.\end{equation}
The term $\Theta^\Tau(\Phi_2)$ guarantees that the $\Phi_2$ point is reached from a genuine QCD splitting of the $\Phi_1$ point. In addition, we require a $1 \to 2$ mapping which preserves $\Tau_0$, ensuring that
\begin{equation} \label{eq:Tau0map}
\Tau_0(\Phi_1^\Tau(\Phi_2)) = \Tau_0(\Phi_2)
\,.\end{equation}
We stress that the use of a $\Tau_0$-preserving mapping ensures that the point-wise singular $\Tau_0$ dependence is alike among all terms in \eqs{1masterfull}{2masterfull} and that the cancellation of said singular terms is guaranteed on an event-by-event basis. In this work, we will use the choice of variable $r_0=\ptja$, the transverse momentum of the hardest jet. A suitable mapping for the fixed order calculation which preserves this quantity was presented in ref.~\cite{Figy:2018imt}; we have, however, constructed our own version, the details of which are provided in \app{splitting_kernels}. 

The nonsingular events arising from nonprojectable regions of $\Phi_1$ and $\Phi_2$ are assigned to the cross sections in \eqs{1belowtau0}{2belowtau1}. The choice of map provides constraints denoted by $\Theta_{\mathrm{map}}$: in the case of the $\Phi_1 \to\widetilde{\Phi}_0$ projection this is the FKS map~\cite{Frixione:2007vw}, while for the $\Phi_2 \to\widetilde{\Phi}_1$ we use our own $\Tau_0$-preserving map. A bar indicates the complement of a Heaviside function.

The term $V_1^C$ denotes the real-virtual contribution, made IR-finite by an appropriate local subtraction,
\begin{align} \label{eq:FOFKS}
  V_1^C(\Phi_1) = V_1(\Phi_1) + \int \frac{\df\Phi_2}{\df \Phi_1^C} \, C_2(\Phi_2)\,,
\end{align}
with $C_2$ a singular approximant of $B_2$. In practice, this is given by FKS subtraction counterterms integrated over the radiation variables
$\df\Phi_2 / \df \Phi_1^C$ and using the singular limit $C$ of the phase space mapping.

In the formul\ae ~involving one or two extra emissions, $U_1$ is a next-to-leading-logarithmic (NLL) Sudakov factor which resums large logarithms of $\Tau_1$, and $U_1'$ its derivative with respect to $\Tau_1$; the $\ord{\alpha_s}$
expansions of these quantities are denoted by $U_1^\one$ and $U_1^{\one\prime}$ respectively.

Finally, the resummation is spread to higher multiplicity phase spaces using normalised splitting probabilities $\mathcal{P}_{N \to N+1}(\Phi_{N+1})$ which satisfy, for every function $f\<\L(\Phi_N,\Tau_N\R)$,
\begin{equation}
\label{eq:cPnorm}
  \int \frac{\d \Phi_{N+1}}{\d \Phi_N} \> f\<\L(\Phi_N,\Tau_N\R) \mathcal{P}_{N
    \to N+1}\<\L(\Phi_{N+1}\R) = \int \d \Tau_N \>
  f\<\L(\Phi_N,\Tau_N\R).
\end{equation}
The radiation phase space is parameterised by $r_N$ and two additional variables, an energy ratio $z$ and an azimuthal angle $\phi$. We discuss the exact functional forms of the $\cP_{N\to N+1}(\Phi_{N+1})$ in more detail in \app{mapping}.

\subsection{Extension of the G{\scriptsize ENEVA} formul\ae~to
  NLL$_{r_1}'$ accuracy}
\label{sec:NLLp1}
We present here an extension to the usual \Geneva~formul\ae~with the
aim of having a better control over the resummation of the transverse
momentum of the second hardest jet. The use of an NLL Sudakov factor in \eqs{1masterfull}{2masterfull} fails to capture all singular terms in the $r_1 \to 0$ limit, in particular those which are proportional to higher order splitting kernels convolved with the PDFs. Upgrading the accuracy of the $r_1$ resummation to NLL$'$, however, remedies this.
Since the formul\ae~are general, we will continue to use the symbols $\Tau_0$ and $\Tau_1$ to identify the 0- and 1-jet resummation variables. 
In order to write the expressions in a more compact way, we introduce
\begin{eqnarray}
  \frac{\d \sigma^{\NLO_0}}{\d \Phi_0}\<\L(\Tau_0^\cut\R) \EI &=&
  \EI B_0\<\L(\Phi_0\R) + V_0\<\L(\Phi_0\R) + \int \frac{\d \Phi_1}{\d
    \Phi_0} \> B_1\<\L(\Phi_1\R) \theta\<\L(\Tau_0<\Tau_0^\cut\R)
  \\
  \frac{\d \sigma^{\NLO_1}}{\d \Phi_1}\<\L(\Tau_1^\cut\R) \EI &=&
  \EI B_1\<\L(\Phi_1\R) + V_1\<\L(\Phi_1\R) + \int \frac{\d \Phi_2}{\d
    \Phi_1} \> B_2\<\L(\Phi_2\R) \theta\<\L(\Tau_1<\Tau_1^\cut \R)
  \\
  \frac{\d \sigma^{\LO_2}}{\d \Phi_2} \EI &=& \EI
  B_2\<\L(\Phi_2\R).
\end{eqnarray}
We also label the contributions from configurations that are not projectable over the underlying phase space (i.e. \eqs{1belowtau0}{2belowtau1}) with a subscript `$\nonproj$'.

At this point, we can define the 0-jet exclusive, 1-jet exclusive and
2-jet inclusive Monte Carlo differential cross sections as
\begin{eqnarray}
  \frac{\d \sigma_0^\MC}{\d \Phi_0}\<\L(\Tau_0^\cut\R) \EI &=& \EI
  \frac{\d \sigma^{\NNLLprime_{\Tau_0}}}{\d
    \Phi_0}\<\L(\Tau_0^\cut\R) - \L. \frac{\d
    \sigma^{\NNLLprime_{\Tau_0}}}{\d \Phi_0}\<\L(\Tau_0^\cut\R)
  \R|_{\NLO_0} + \frac{\d \sigma^{\NLO_0}}{\d
    \Phi_0}\<\L(\Tau_0^\cut\R)
  \\
  \frac{\d \sigma_1^\MC}{\d \Phi_1}\<\L(\Tau_1^\cut\R) \EI &=& \EI
  \L\{\L[\frac{\d \sigma^{\NNLLprime_{\Tau_0}}}{\d \Phi_0 \> \d
      \Tau_0} - \L. \frac{\d \sigma^{\NNLLprime_{\Tau_0}}}{\d
      \Phi_0 \> \d \Tau_0} \R|_{\NLO_1}\R] \mathcal{P}_{0 \to 1}\<\L(\Phi_1\R)
  U_1\<\L(\Phi_1,\Tau_1^\cut\R) \R.
  \nonumber \\
  && \EI \L. {} + \frac{\d \sigma^{\NLO_1}}{\d
    \Phi_1}\<\L(\Tau_1^\cut\R) + \frac{\d
    \sigma^{\NLLprime_{\Tau_1}}}{\d \Phi_1}\<\L(\Tau_1^\cut\R) -
  \L. \frac{\d \sigma^{\NLLprime_{\Tau_1}}}{\d
    \Phi_1}\<\L(\Tau_1^\cut\R) \R|_{\NLO_1}\R\} \theta\<\L(\Tau_0 >
  \Tau_0^\cut\R)
  \nonumber \\
  && \EI {} + \frac{\d \sigma_{\nonproj}^{\LO_1}}{\d \Phi_1} \>
  \theta\<\L(\Tau_0 < \Tau_0^\cut\R)
  \\
  \frac{\d \sigma_2^\MC}{\d \Phi_2} \EI &=& \EI \L\{\L[\frac{\d
      \sigma^{\NNLLprime_{\Tau_0}}}{\d \Phi_0 \> \d \Tau_0} -
    \L. \frac{\d \sigma^{\NNLLprime_{\Tau_0}}}{\d \Phi_0 \> \d
      \Tau_0} \R|_{\NLO_1}\R] \mathcal{P}_{0 \to 1}\<\L(\Phi_1\R)
  U_1'\<\L(\Phi_1,\Tau_1\R) \mathcal{P}_{1 \to 2}\<\L(\Phi_2\R) \R.
  \nonumber \\
  && \EI \L. {} + \frac{\d \sigma^{\LO_2}}{\d \Phi_2} +
  \L[\frac{\d \sigma^{\NLLprime_{\Tau_1}}}{\d \Phi_1 \> \d
      \Tau_1} - \L. \frac{\d \sigma^{\NLLprime_{\Tau_1}}}{\d
      \Phi_1 \> \d \Tau_1} \R|_{\LO_2}\R] \mathcal{P}_{1 \to
    2}\<\L(\Phi_2\R)\R\} \theta\<\L(\Tau_1 > \Tau_1^\cut\R)
  \theta\<\L(\Tau_0 > \Tau_0^\cut\R)
  \nonumber \\
  && \EI {} + \frac{\d \sigma_{\nonproj}^{\LO_2}}{\d \Phi_2} \>
  \theta\<\L(\Tau_1 < \Tau_1^\cut \R) \theta\<\L(\Tau_0 >
  \Tau_0^\cut\R).
  \label{eq:NLLprimesigma2}
\end{eqnarray}
The above formul\ae~still reproduce the exact
$\NNLLprime$ spectrum of the $\Tau_0$ resummation. In order to prove
this, we note that, since the resummed and resummed expanded terms are
built such that they have the same total cumulant,
\begin{eqnarray}
  && \EI \frac{\d \sigma^{\NLLprime_{\Tau_1}}}{\d
    \Phi_1}\<\L(\Tau_1^\cut\R) - \L. \frac{\d
    \sigma^{\NLLprime_{\Tau_1}}}{\d \Phi_1}\<\L(\Tau_1^\cut\R)
  \R|_{\NLO_1}
  \nonumber \\
  && {} + \int \frac{\d \Phi_2}{\d \Phi_1} \L[\frac{\d
    \sigma^{\NLLprime_{\Tau_1}}}{\d \Phi_1 \> \d \Tau_1} -
  \L. \frac{\d \sigma^{\NLLprime_{\Tau_1}}}{\d \Phi_1 \> \d \Tau_1}
  \R|_{\LO_2}\R] \mathcal{P}_{1 \to 2}\<\L(\Phi_2\R) \theta\<\L(\Tau_1
> \Tau_1^\cut\R) = 0
\end{eqnarray}
and the total cumulant of the Sudakov form factor must be
\begin{equation}
  U_1\<\L(\Phi_0,\Tau_1^\cut\R) + \int \frac{\d \Phi_2}{\d \Phi_1} \>
  U_1'\<\L(\Phi_0,\Tau_1\R) \mathcal{P}_{1 \to 2}\<\L(\Phi_2\R)
  \theta\<\L(\Tau_1 > \Tau_1^\cut\R) = 1.
\end{equation}
Using the above formulae to integrate over $\Tau_1$, we can write the
1-jet inclusive differential cross section as
\begin{eqnarray}
  \frac{\d \sigma_{\geq 1}^\MC}{\d \Phi_1}\<\L(\Tau_1^\MAX\R) \EI &=&
  \EI \L\{\L[\frac{\d \sigma^{\NNLLprime_{\Tau_0}}}{\d \Phi_0 \>
      \d \Tau_0} - \L. \frac{\d \sigma^{\NNLLprime_{\Tau_0}}}{\d
      \Phi_0 \> \d \Tau_0} \R|_{\NLO_1}\R] \mathcal{P}_{0 \to 1}\<\L(\Phi_1\R) +
  \frac{\d \sigma^{\NLO_1}}{\d \Phi_1}\<\L(\Tau_1^\MAX\R)\R\}
  \theta\<\L(\Tau_0 > \Tau_0^\cut\R)
  \nonumber \\
  && \EI {} + \frac{\d \sigma_{\nonproj}^{\LO_1}}{\d \Phi_1} \>
  \theta\<\L(\Tau_0 < \Tau_0^\cut\R),
\end{eqnarray}
which proves that the $\NNLLprime$ accuracy of the $\Tau_0$
resummation at partonic level is preserved.

\subsection{Jet veto resummation for colour singlet$+0-$jet at NNLL$'$}
\label{sec:0jet}
The factorisation of the $0-$jet cross section for colour singlet production with a jet algorithm was first achieved in refs.\cite{Becher:2012qa,Tackmann:2012bt,Stewart:2013faa}; expressed in the language of the rapidity renormalisation group~\cite{Chiu:2011qc,Chiu:2012ir} and for an anti-$k_T$ jet~\cite{Cacciari:2008gp} of radius $R$, the cumulant up to $\ptv$ reads
\begin{align}
\frac{\mathrm{d}\sigma}{\mathrm{d}\Phi_0}(\ptv,\mu,\nu) \< = \<
\sum_{a,b} H_{ab}(\Phi_0,\mu) B_a(Q,\ptv,R,x_a,\mu,\nu)
B_b(Q,\ptv,R,x_b,\mu,\nu) S_{ab}(\ptv,R,\mu,\nu)
\label{eq:factorisation}
\end{align}
The right hand side of \eq{factorisation} features hard, soft and beam functions which describe high energy ($\sim Q$), isotropic soft and collinear radiation respectively. While the hard function is a process-dependent object, both soft and beam functions are universal and depend only upon the flavour of the initiating partons. 

The hard function arises from the matching of QCD onto SCET, and as such is simply given by the squared amplitude of the process under consideration. The beam functions, on the other hand, are intrinsically nonperturbative objects; for $\ptv\gg \lqcd$, however, it is possible to perform an operator product expansion (OPE) and to write them as
\begin{align}
B_i(Q,\ptv,R,x,\mu_B,\nu_B)=\sum_j \int_x^1\frac{\mathrm{d}\xi}{\xi}\mathcal{I}_{ij}(Q,\ptv,R,\xi,\mu_B,\nu_B)\,f_j\left(\frac{x}{\xi},\mu_B\right)\,,
\end{align}
where the $\mathcal{I}_{ij}$ are perturbatively calculable matching coefficients which are convolved with the standard collinear PDFs $f_j$. Finally, the soft function is also a nonperturbative object and is defined operatorially as a forward scattering matrix element of soft Wilson lines along the incoming beam directions. It also satisfies an OPE, meaning that we are able to treat it as a perturbatively calculable object and neglect the nonperturbative component. This latter contribution is delegated to the hadronisation model included in the parton shower to which we interface. 

We note that the structure of \eq{factorisation} is rather different from the factorisation theorems that have been used in previous \geneva implementations, in several respects. First, the nature of the observable means that the hard, beam and soft functions are combined multiplicatively, rather than via a convolution as would be the case for the $N-$jettiness observable. Second, the observable requires a SCET-II treatment, which introduces the rapidity scale $\nu$ into the problem. This is necessary to distinguish between soft and collinear modes in the Lagrangian, which share a common invariant mass scaling. The soft and beam functions each contain divergences in a rapidity regulator $\alpha$, which however cancel in the cross section, leaving behind rapidity logarithms.\footnote{We use the exponential regulator of ref.~\cite{Li:2016axz}.} Third, it is convenient to write the factorisation formula in the cumulant of the observable $\ptv$ rather than in the spectrum, i.e. differentially. Each of these facts entails minor modifications to the partonic calculation in \geneva with respect to the jettiness case, which we detail in \sec{details}.

\Eq{factorisation} features logarithms of the scale ratios $Q/\mu$,
$\mu/\ptv$ and $\nu/\ptv$, each of which could potentially be large
and spoil the convergence of an expansion in $\alpha_s$. This is
avoided by judicious choices of separate virtuality scales $\mu_H$,
$\mu_B$, $\mu_S$ and rapidity scales $\nu_B$, $\nu_S$ which minimise the size of the logarithms in each component. Evolution to common scales $\mu$, $\nu$ is then achieved via the renormalisation group, which serves to resum the large logarithms. The resummed expression can then be written as
\begin{align}
\frac{\mathrm{d}\sigma^{\mathrm{resum}}}{\mathrm{d}\Phi_0}(\ptv,\mu,\nu)&=\sum_{a,b}H_{ab}(\Phi_0,\mu_H)\,B_a(Q,\ptv,R,x_a,\mu_B,\nu_B)\,B_b(Q,\ptv,R,x_b,\mu_B,\nu_B)\nn\\
&\qquad\qquad \times\,S_{ab}(\ptv,R,\mu_S,\nu_S)\,U(\mu,\nu;\mu_H,\mu_B,\mu_S,\nu_B,\nu_S)
\label{eq:resummation}
\end{align}
where the evolution function $U$ is a product of factors for each of the hard, beam and soft sectors, each given by the exponential of the relevant anomalous dimension. Resummation at NNLL$'$ accuracy requires knowledge of each of the hard, soft and beam functions up to two-loop order. The relevant soft function has been calculated in refs.~\cite{Stewart:2013faa,Abreu:2022sdc}, while the beam function calculations appear in refs.~\cite{Stewart:2013faa, Bell:2022nrj, Abreu:2022zgo}. In addition, the cusp (noncusp) anomalous dimensions are required to three (two) loops; the expressions are readily available in the literature~\cite{Moch:2004pa,Vogt:2004mw,Korchemsky:1987wg,Tarasov:1980au,Larin:1993tp,Stewart:2013faa}.

In ref.~\cite{Tackmann:2012bt}, it was shown that placing a jet veto can mix the phase space constraints on the different sectors of the SCET Lagrangian, leading to the presence of soft-collinear mixing terms. Various ways for treating these terms have been proposed in the literature -- given that in this work we will use the results of refs.~\cite{Abreu:2022zgo} and~\cite{Abreu:2022sdc} for the soft and beam functions, we follow their prescription and exponentiate the mixing terms rather than treating them as an additional contribution at fixed order.

We do not consider the resummation of logarithms of the vetoed jet radius $R_v$\footnote{We distinguish here between the hard jet radius $R_J$ and the radius of vetoed jets below $\ptvb$, $R_v$. Although these have different physical significance, in practice we always take $R_J=R_v=R$.}, which was achieved via a numerical approach in refs.~\cite{Dasgupta:2014yra,Banfi:2015pju}. The effect of this resummation is expected to be small for experimental values of the jet radius 0.4-0.5, but could in principle be included in our formalism by modifying the expression for the rapidity anomalous dimension to include the known numerical terms. In addition, we do not consider the effect of placing rapidity cuts on the jets, which are often experimentally necessary to reduce pileup effects and because of limited detector acceptance. The resummation formalism in the presence of such cuts has been developed in ref.~\cite{Michel:2018hui} and the resummation achieved up to NLL$'$. In ref.~\cite{Campbell:2023cha}, it was shown that for the $WW$ process, a cut of $y_{\cut}=2.5$ can have an effect of a few percent on the exclusive cross section, while for $y_{\cut}=4.5$ the effect is negligible. Since the experimental analyses which we consider in \sec{expdata} make use of looser cuts $\sim 4.5$, our omission is justified in this case.
 
\subsection{Jet veto resummation for colour singlet$+1-$jet at NLL$'$}
\label{sec:1jet}
The factorisation for the production of a colour singlet in association with a hard jet, vetoing softer jets below a threshold $\ptv$, was first studied in ref.~\cite{Liu:2012sz}. An extension of this work appeared in ref.~\cite{Liu:2013hba}, where the ingredients necessary to reach NLL$'$ accuracy were provided. In this section, we base our discussion on ref.~\cite{higgsjet}, where the factorisation for this class of processes and this observable was revisited. The factorisation formula reads
\begin{align}\label{eq:factorisation1jet}
\frac{\mathrm{d} \sigma }{\mathrm{d} \Phi_1} (\ptv,\mu,\nu)  = \sum_{\kappa} & H_{\kappa} (\Phi_1, \mu) B_a (Q,\ptv,R, x_a, \mu, \nu)\,  B_b(Q,\ptv, R, x_b, \mu, \nu) S_{\kappa}(\ptv, y_J, \mu, \nu) \, \nn \\
&\times \SC_j(\ptv R, \mu)\,  J_j(p_T^J R, \mu) \SCNG_j\bigg(\frac{\ptv}{p_T^J}\bigg)\,.
\end{align}
In general, the production of a colour singlet in association with an additional parton may proceed via several possible channels -- we therefore use the subscript $\kappa$ to denote the set of partons $\{a,b,j\}$ which make up the initial states and the jet respectively. The collinear sector of this expression consists of beam functions, which are identical to those in the zero-jet case, and an additional jet function to describe final-state collinear radiation. The soft sector has been refactorised into components describing global soft and soft-collinear radiation, and a nonglobal term. The first describes isotropic soft radiation which has no information about the radius of the hard jet $R_J$, while the second describes soft emission from a single Wilson line with colour factor $C_j$ which probes the jet boundary. The third instead describes logarithms of $\ptv/p_T^J$ which arise due to the phase space constraints placed by the finite jet size. The full details of the factorisation proof and the expressions for the various terms in this formula will be provided in an upcoming work~\cite{higgsjet}.

Resummation proceeds along the same lines as in the $0-$jet case. At NLL accuracy, the resummed cross section is simply given by a product of the leading order hard function (i.e. the Born matrix element) multiplying an evolution factor, with cusp (noncusp) anomalous dimensions included at two (one) loops. The $U$ factor appearing in e.g. \eq{1masterfull} is then simply given by
\begin{align}
U^{\mathrm{NLL}}=\prod_{i\in H,B,S,J,\SC,\SCNG} U_i^{\mathrm{NLL}}
\end{align}
where the index $i$ runs over each sector. In order to reach NLL$'_{r_1}$ accuracy, the boundary terms are also required at one-loop order. The jet functions can be found in refs.~\cite{Liu:2012sz,Arratia:2020nxw}, while the expressions for the global soft and soft-collinear functions appear in refs.~\cite{Liu:2013hba,higgsjet}. We adopt the same treatment of nonglobal logarithms as in ref.~\cite{higgsjet}, including the 5-loop expanded solution of the Banfi-Marchesini-Smye equation~\cite{Banfi:2002hw} which resums these terms at LL accuracy~\cite{Schwartz:2014wha}. We stress that we do not claim that $1-$jet exclusive cross section will be NLL$'$ accurate in all kinematic regimes~\footnote{This should, however, be true at parton level for large values of $\ptja$.}; rather, we include the resummation only as a tool to facilitate the separation of the $1-$ and $2-$jet bins. 

\section{Technical details of the calculation}
In this section we provide details about our implementation of the formul\ae~presented in \sec{theory}. We explain our interface to the resummed calculation in \scetlib, the procedures adopted for switching off the resummation and estimating theoretical uncertainties and our interface to the parton shower.
\label{sec:details}
\subsection{\scetlib interface}
Our implementations of \eq{resummation} and \eq{factorisation1jet}
rely on the \texttt{C++} library \scetlib~\cite{scetlib}, which
facilitates numerical calculations in soft-collinear effective
theory. We have augmented the existing NNLL$'$ implementation of the
jet veto resummation for colour-singlet production in the gluon fusion
channel~\cite{Stewart:2013faa} with the results of
refs.~\cite{Abreu:2022sdc, Abreu:2022zgo}. These provide the full $R$
dependence in the soft and beam functions, as well as the missing
two-loop term in the quark beam function proportional to
$\delta(\ptv)$. We are then able to call the \scetlib functions
providing the process-independent part of the resummed calculation
directly from \textsc{Geneva}, and combine this information with the
hard functions which have been implemented in the \geneva code. We note that, in contrast to $N$-jettiness resummation, in \eq{factorisation} it is the cumulant of the distribution and not the spectrum that is written in terms of hard, soft and beam functions. In order to obtain the resummed spectrum differential in $\ptja$, we differentiate the cumulant numerically using functionality provided by the \texttt{gsl} library. 

\subsection{Process-specific ingredients}
Although the theoretical framework we have described in \sec{theory} could be applied to any colour-singlet production process (with minor modifications required in cases where final-state photons are present~\cite{Alioli:2020qrd,Cridge:2021hfr}), in this work we restrict our attention to the diboson process $pp \to W^+W^-$ with different flavour leptons in the final state. In order to avoid contamination from the $t\bar{t}$ process, we work in a four-flavour scheme in which the $b$-quark is not considered a constituent of the proton and we correspondingly run $\alpha_s$ with only four flavours. We take the necessary tree and one-loop amplitudes from \openloops\cite{Buccioni:2019sur}. We also require the two-loop hard function, which we construct from the squared amplitudes first computed in ref.~\cite{Gehrmann:2015ora} and made available in the \texttt{VVAmp} package. Since the two-loop matrix elements are currently only available for massless quark loops, we neglect any two-loop Feynman
diagram where bottom or top quarks appear in a loop.

In addition to the $q\bar{q}$ channel, a gluon-initiated, loop-induced channel becomes available starting at NNLO. We include this contribution at NLL, again taking the one-loop squared matrix elements from \texttt{OpenLoops}, which also include the contribution from an off-shell Higgs boson decaying to a $WW$ pair. For small values of the jet veto scale, the relative size of the $gg$-initiated channel is negligible -- it increases to around $6-8\%$ percent of the total cross section, however, for veto scales $\sim 60~\GeV$~\cite{Campbell:2023cha}. 

We work in the approximation of a diagonal CKM matrix $V$. We note
that, as long as the first two generations of quarks are treated as
massless, the results where two $W$ bosons are attached to a single
quark line would be equivalent if we worked with a block-diagonal
matrix where $V_{us}$ and $V_{cd}$ are allowed to be non-zero. Indeed,
attaching the two $W$ bosons to a quark line that connects the
flavours $\alpha$ and $\beta$ and summing over the flavours $\gamma$
of the intermediate quarks, provided they are massless, gives the
factor
\begin{equation}
  \sum_k V^*_{\alpha\gamma} V_{\beta\gamma} = \sum_k V_{\gamma\alpha}
  V^*_{\gamma\beta} = \delta_{\alpha\beta},
\end{equation}
where the equivalence is guaranteed by the unitarity of $V$. The only
non-diagonal terms of the CKM matrix that can affect the final result
are those in which non-zero masses cause intermediate flavours to have
different propagators. The size of these effects is further limited by
the fact that the massive quarks can only appear as internal
propagators and not as external legs, meaning that each squared matrix
element carrying non-diagonal effects will be proportional to the
squared modulus of two of the four elements $V_{ub}$, $V_{cb}$,
$V_{td}$ and $V_{ts}$. The largest possible effect will then be
proportional to the fourth power of the largest of the moduli, which
is $\L|V_{cb}\R|^4 \sim 2.8 \times 10^{-6}$. The case where the two
$W$ bosons are attached to different quark lines instead only appears
at order $\alphaS^2$ and the size of its effect can be estimated to be
proportional to $\alphaS^2 \L|V_{us}\R|^2 \sim 7.0 \times
10^{-6}$. Both numbers are lower than the theoretical precision we
aim to achieve.

\subsection{Profile scales and theoretical uncertainties}
\label{sec:profile_scales_theoretical_uncertainties}

For values of $\ptja$ near the hard scale $\sim Q$, the factorisation formula in \eq{factorisation} breaks down and the fixed order calculation is required to provide a correct theoretical description. We must therefore provide a prescription to turn off the resummation before the exponentiated singular terms become too large. This can be achieved in a smooth manner by employing profile scales, first introduced in refs.~\cite{Ligeti:2008ac,Abbate:2010xh,Berger:2010xi}. These profiles evolve the beam and soft scales to the hard scale as a function of $\ptva$ and hence stop the RG evolution and resummation when the common scale $\mu_{\rm{NS}} = \mu_S = \mu_B = \mu_H$ is reached. Though there is some freedom in their exact definition, we follow the conventions adopted by e.g. ref.~\cite{Stewart:2013faa} and use the forms
\begin{align} \label{eq:centralscale}
\mu_H &= \mu_{\rm{NS}}\nn\,,  \\
\mu_B &= \mu_S = \nu_S = \mu_\NS\ f_{\rm run}\left(\ptva/Q\right)\,, \\
\nu_B &= \mu_{\rm{NS}} \,,\nn
\end{align}
where the common profile function $f_{\rm run}(x)$ is given by~\cite{Stewart:2013faa}
\begin{align}
f_{\rm run}(x) &=
\begin{cases} x_0 \bigl[1+ (x/x_0)^2/4 \bigr] & x \le 2x_0\,,
\\ x & 2x_0 \le x \le x_1\,,
\\ x + \frac{(2-x_2-x_3)(x-x_1)^2}{2(x_2-x_1)(x_3-x_1)} & x_1 \le x \le x_2\,,
\\  1 - \frac{(2-x_1-x_2)(x-x_3)^2}{2(x_3-x_1)(x_3-x_2)} & x_2 \le x \le x_3\,,
\\ 1 & x_3 \le x\,.
\end{cases}
\label{eq:frun}
\end{align}
This choice ensures that the resummation is switched off above $x_3$, while for $x<x_1$ the scales follow the canonical values dictated by the RGE,
\begin{align}
&\mu_H\sim Q, \qquad \mu_B\sim \mu_S\sim \ptva,\nn\\
&\nu_B\sim Q, \qquad \nu_S\sim \ptva,
\end{align}
which minimise the size of the logarithms in \eq{factorisation}. Between $x_1$ and $x_3$ a smooth transition between the resummation and fixed order regions is ensured, while below $x_0$ the scales asymptote to a fixed value, preventing $\alpha_s$ from being evaluated at a nonperturbative scale.

\begin{figure}[t]
   \centering
   \includegraphics[width=\rescaletwoplots]{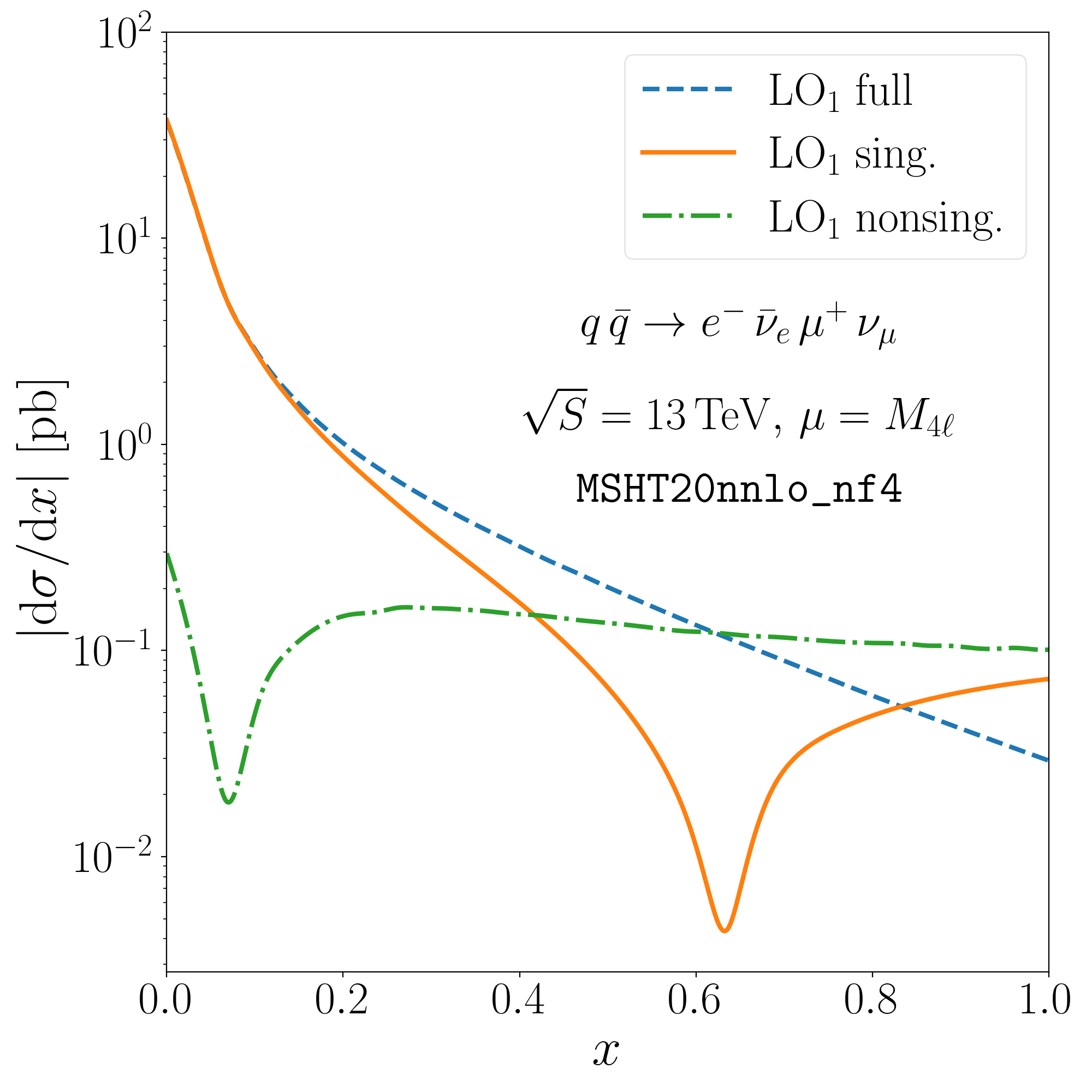}%
   \includegraphics[width=\rescaletwoplots]{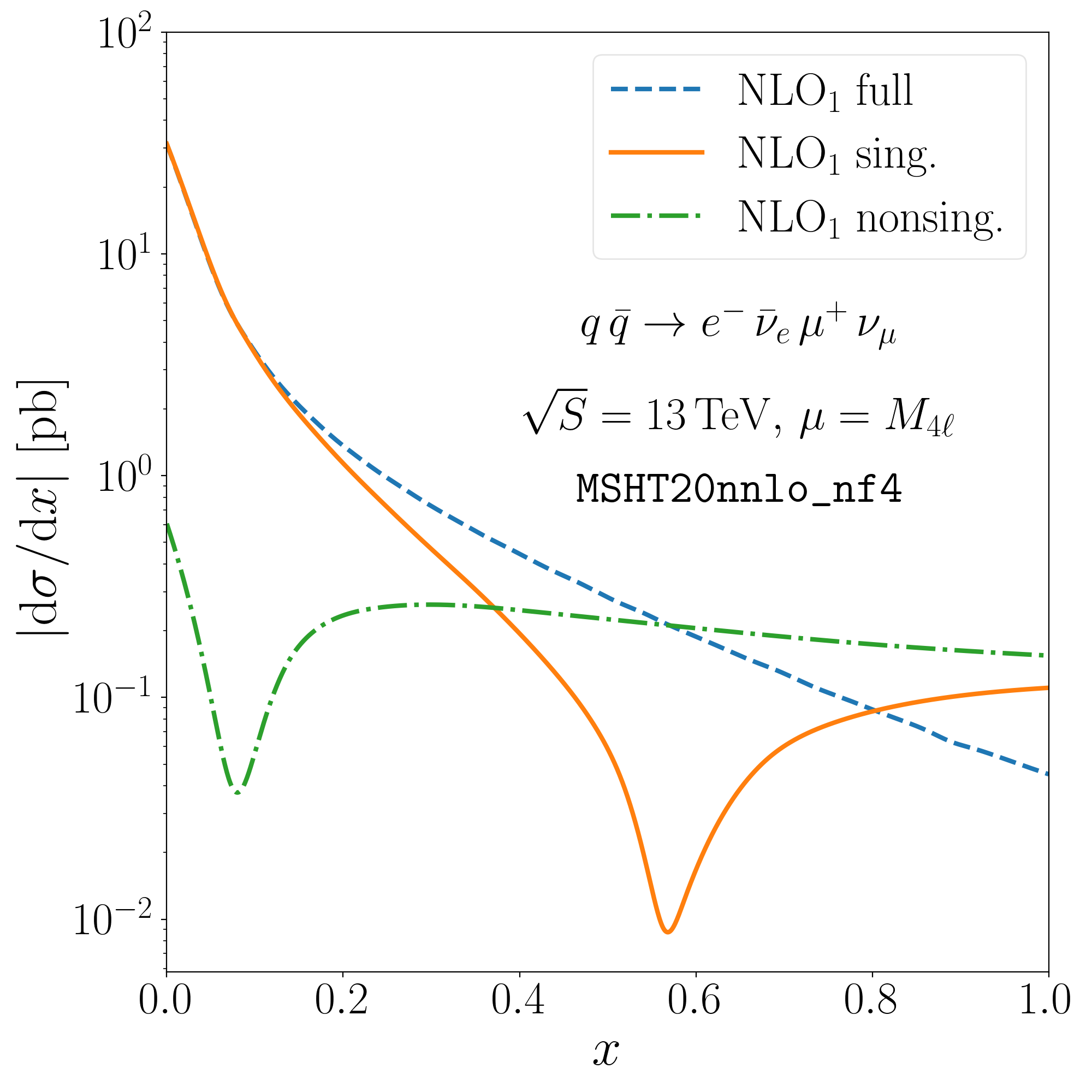}%
   \caption{Comparison of the fixed-order, singular, and nonsingular spectra as a function of $x=\ptja/Q$ for $pp\to W^+W^- \to e^-\bar{\nu}_e\mu^+\nu_\mu$ at LO$_1$ (left) and at NLO$_1$ (right).}
   \label{fig:choosetrans}
\end{figure}

Our choice of the transition points $x_i$ is guided by an examination of the relative sizes of the singular and nonsingular contributions to the cross section as a function of $x$. In \fig{choosetrans} we plot these contributions at NLO (i.e., spectrum at LO$_1$) and at NNLO (spectrum at NLO$_1$). We note that at both orders, the singular and nonsingular contributions become of approximately equal size near $x=0.4$ and therefore require the approach to the fixed order to begin around this point. For our central profiles, we make the choice
\begin{align}
x_0=2.5~\GeV/Q,\qquad \{x_1,x_2,x_3\}=\{0.15,0.4,0.6\}
\end{align}
where $Q=M_{WW}$, which coincides with that made in ref.~\cite{Stewart:2013faa}.

We estimate the uncertainties associated with our scale choices by performing various kinds of variation. Our fixed-order uncertainties are obtained by varying $\mu_{\rm NS}$ up and down by a factor of two, thus preserving all scale ratios appearing inside the logarithms. We also vary our transition points $x_i$ collectively by an amount $\pm 0.05$ for our central choice of $\mu_{\rm NS}$ to gauge the uncertainty associated with this choice. Our resummation uncertainties are obtained by fixing $\mu_H$ and multiplying the individual beam and soft scales by a variation function,
\begin{align}
\rho_i^{\uparrow\downarrow}\left(\ptva\right)=\rho_i^{\rm central}\left(\ptva\right) \left[f_{\rm vary}\left(\ptva/Q\right)\right]^{\pm 1}
\end{align}
where $i\in \{B,S\}$, $\rho \in \{\mu,\nu\}$ and we have 
\begin{align}
  f_{\mathrm{vary}}(x)=\begin{cases}
                    2(1-x^2/x_3^2) &  0\leq x \leq x_3/2  \\
                    1+2(1-x/x_3)^2 & x_3/2\leq x\leq x_3  \\
                    1 & x_3\leq x\,.
  \end{cases}\,
\end{align}
This form of $f_{\rm vary}$ ensures that the resummation variations are turned off smoothly for $\ptva\sim Q$. We consider the subset of possible variations of $\{\mu_S, \nu_S, \mu_B, \nu_B\}$ for which the arguments of the logarithms 
\begin{align}
\frac{\mu_S}{\mu_B}\sim \frac{\mu_S}{\nu_S}\sim 1, \qquad \frac{\nu_B}{\nu_S}\sim \frac{Q}{\ptva}
\end{align}
are varied by no more than a factor of two in the resummation region (and thus exclude possibilities such as $\{\nu_B^{\uparrow},\nu_S^{\downarrow}\}$).

In total, we are left with a total of 40 different variations: two associated with the fixed-order variations, two associated with the transition point variations and 36 associated with the resummation. We combine these quantities differently depending on the observable we consider. For inclusive quantities, such as the rapidity of the colour singlet system, we simply envelope the fixed-order variations. For the exclusive observables of interest, the exact construction of the uncertainty relies on a correct propagation of the individual sources from each jet multiplicity bin, which can be described using a covariance matrix~\cite{Stewart:2013faa}. We will primarily be interested in the exclusive $0-$jet cross section, for which (following ref.~\cite{Stewart:2013faa}) we define a yield uncertainty,
\begin{align}
\Delta_{\mu 0}\left(\ptva\right) = \max_{v_i\in V_\mu}\left|\sigma_0^{v_i}\left(\ptva\right)-\sigma_0^{\rm central}\left(\ptva\right)\right|
\label{eq:yieldunc}
\end{align}
where $V_\mu$ runs over the transition point and fixed-order variations, and a resummation uncertainty
\begin{align}
\Delta_{\rm resum}\left(\ptva\right)=\max_{v_i\in V_{\rm resum}} \left|\sigma_0^{v_i}\left(\ptva\right)-\sigma_0^{\rm central}\left(\ptva\right)\right|\,,
\label{eq:resumunc}
\end{align}
where $V_{\rm resum}$ denotes the 36 resummation variations. We then combine these in quadrature to obtain our final estimate.

Having ensured that the $\ptva$ resummation is correctly switched off in the fixed-order region, we can apply the same technique to the $\ptvb$ resummation. In this case, following ref.~\cite{higgsjet}, we run all scales to a common hard scale $\mu_H$ and employ hybrid profile scales which take the form
\begin{align}
 g_\text{run}(\xi; \mu_0, \mu_H )=   h_\text{run}(\xi) \, \mu_0    +  [1-h_\text{run}(\xi)] \, \mu_H,
\end{align}
with
\begin{align} 
h_\text{run}(\xi) &= \begin{cases}
1 & 0 < \xi \leq \xi_1 \,, \\
1- \frac{(\xi-\xi_1)^2}{(\xi_2-\xi_1)(\xi_3-\xi_1)} & \xi_1 < \xi \leq \xi_2
\,, \\
\frac{(\xi-\xi_3)^2}{(\xi_3-\xi_1)(\xi_3-\xi_2)} & \xi_2 < \xi \leq \xi_3
\,, \\
0 & \xi_3 \leq \xi
\,.\end{cases}
\end{align}
The profile scales are then given by
\begin{align}
\mu_H&= Q\, ,  \nn \\
\mu_{B} = \mu_S =\nu_S &= g_\text{run}\left(\xi; \, \ptvb,  \mu_H\right)\, , \nn \\
\mu_{J}  &= g_\text{run}\left(\xi; \, \ptja \Rj,  \mu_H\right)\, , \nn \\
\mu_{\mathcal{S}}  &= g_\text{run}\left(\xi; \, \ptvb \Rj,  \mu_H\right)
\,, \nn \\
\nu_{B}&= g_\text{run}(\xi; \, \omega, \, \mu_H), 
\end{align}
where $\omega_{a,b}=M_{VJ}\,e^{\pm Y_{VJ}}$ and $\xi=\ptvb/\ptja$. We stress that although we have denoted the hard, soft and beam scales as $\mu_H$, $\mu_B$ and $\mu_S$ in both the $\ptva$ and $\ptvb$ resummation formul\ae~in order to avoid a proliferation of notation, these have different interpretations and consequently different expressions in the two cases. We choose the values of the transition points to be 
\begin{align}
\{\xi_1,\xi_2,\xi_3\}=\{0.2,0.6,1.0\}\,,
\end{align}
where the choice $\xi_3=1$ ensures that the resummation is fully switched off when $\ptvb=\ptja$. We do not associate any uncertainties with this choice of profiles, since these are expected to be subleading with respect to the uncertainties described by \eqs{yieldunc}{resumunc}.
We have checked, however, that when fixing $\xi_3=1,\,\xi_2=(\xi_1+\xi_3)/2$ and varying $\xi_1$ by 0.2 there is no visible difference in our results. Furthermore, we do not attempt to estimate uncertainties arising from higher-order clustering effects, neither in the $0-$jet nor in the $1-$jet resummation. We justify this omission by noting that the NLO clustering corrections were calculated in ref.~\cite{Alioli:2013hba} and were found to be small.

Before moving on, we remark that a notable difference between this and previous \geneva implementations employing SCET is that the factorisation formula, \eq{factorisation}, is for the cumulant and not the spectrum. Consequently, we set the scales in the cumulant -- this removes the need for a cross section fix term (see e.g. refs.~\cite{Alioli:2015toa,Alioli:2019qzz}), which serves to account for the fact that the operations of scale setting and integration do not commute when the profiles are functions of the resolution variable.

\subsection{Matching to the shower}
\label{sec:shower}
The purpose of the parton shower is to promote the hard, single-parton `jets' created by the resummation to full jets by adding soft and collinear emissions, and to create new jets in the inclusive $2-$jet bin. In order to prevent the shower from double-counting regions of phase space already covered by the resummation, it is important to make careful choices of shower starting scales and to apply vetoing procedures in cases where the shower should not be allowed to emit. We detail our approach in the following, considering each multiplicity of jet bin separately. 

In the case of the $0-$jet bin, the shower should  restore the emissions which were integrated over in the construction of the $0-$jet cross section. We achieve this by setting the starting scale for the $0$-jet events to be equal to $\ptva$ and additionally requiring that the transverse momentum of the hardest jet after the shower does not exceed $\ptva$.\footnote{This can happen even for $p_T$ ordered showers when final-state emissions are considered~\cite{Corke:2010zj}.} We re-shower events which do not fulfil these criteria. We remark that $0-$jet events only account for a small fraction $\mathcal{O}(1\%)$ of the total cross section.

The $1-$ and $2-$jet bins deserve a greater level of discussion. In the original \geneva implementation for hadronic collisions~\cite{Alioli:2015toa}, the mismatch between the resolution variable $\mathcal{T}_0$ and the shower ordering variable meant that it was desirable to reduce the size of the $1-$jet cross section. This was achieved by multiplying the $1-$jet bin by an additional Sudakov form factor $U_1(\ptvb, \Lambda_1)$ (and making corresponding modifications to the $2-$jet bin). In this work, our use of $\ptja$ and $\ptjb$ as resolution variables removes the need for this procedure, since these are more closely related to the ordering variable of the shower. In addition, the use of NLL$'$ rather than NLL resummation for $\ptjb$ ensures that the contribution in \eq{NLLprimesigma2} is truly nonsingular in the $\ptjb\to 0$ limit, meaning that these events can be safely passed to the shower.

We choose the shower starting scale to be equal to $\ptvb$ for the $1-$jet events and to the $p_T$ of the second hardest parton in the event, i.e. $\ptjb$, for the $2-$jet events. We then allow the shower to run and veto after each emission if either of the following conditions are satisfied:
\begin{enumerate}
\item For initial-state emissions, the transverse momentum of the emitted parton with respect to the beam $k_T>\ptjb$.
\item For final-state emissions, the emitted and sister partons do not belong to the same jet and each have a transverse momentum with respect to the beam axis $k_T>\ptjb$.
\end{enumerate}
After completion of showering, we also check that the third hardest
jet has a transverse momentum less than the starting scale. In this
way, we ensure that the two hardest jets in the event originate from
the two hardest partons created by the resummation and that we do not
double count any region of phase space by showering.

In refs.~\cite{Alioli:2015toa,Alioli:2022dkj}, it was demonstrated that the NNLL$'$ accuracy of the $\mathcal{T}_0$ variable resummed in those works was affected by the shower only at the N$^3$LL level. In ref.~\cite{Alioli:2021qbf}, where instead the colour-singlet transverse momentum was used as a resolution variable and was resummed at N$^3$LL accuracy, no such claim was made: nonetheless, it was possible to show that, by using the \texttt{dipoleRecoil} setting of the \pythia parton shower, the accuracy of the $q_T$ distribution was numerically preserved by the shower to within the Monte Carlo accuracy of the calculation. In our current context, our resolution variable (the jet transverse momentum) is by its nature extremely sensitive to the pattern of soft and collinear emissions generated by the shower. We do not, therefore, claim any formal logarithmic accuracy for the distribution after showering beyond that which is provided by the shower itself. This is due solely to final-state emissions; in the initial-state case, the ordering variable of the shower is very closely related to our resolution variable and we have checked numerically that the observable is perfectly preserved by the shower. We shall examine the numerical size of the differences induced by the shower on the partonic results (which are formally NNLL$'$+NNLO accurate) in \sec{showeredresults}.

\section{Results}
\label{sec:results}
In this section we present our numerical results. We begin by
validating the NNLO accuracy of our partonic calculation against the
fixed order code \Matrix, before comparing our resummed predictions
against a similar implementation in MCFM. We then show a comparison of
partonic results with those after interfacing to the \pythia parton
shower, and compare our final predictions with data collected from the
ATLAS and CMS experiments.

\subsection{Physical parameters}

All the results of this paper are obtained in the $\nf = 4$ scheme,
where only the first two generations of quarks are treated as
massless. The bottom and top masses are instead set to
\begin{equation}
  \mb = 4.18 \ \GeV
  \qquad
  \mt = 173.1 \ \GeV.
\end{equation}
If not stated otherwise, we use the four-flavour PDF set
\verb|MSHT20nnlo_nf4|~\cite{Bailey:2020ooq,Cridge:2021qfd} from
\LHAPDF~\cite{Buckley:2014ana}.

The electroweak constants are set in the $G_\mu$ scheme, where the
Fermi constant $G_\mu$ and the masses $\mW$ and $\mZ$ and widths
$\GammaW$ and $\GammaZ$ of the $W$ and $Z$ bosons are taken as
independent parameters, from which the electroweak coupling $\alphaEW$
and the Weinberg angle $\thetaW$ are derived. Furthermore, we work in
the complex-mass scheme~\cite{Denner:2005fg} and define the complex
masses of the $V = W, Z$ bosons as
\begin{equation}
  \muV^2 = \mV^2 - i\GammaV\mV.
\end{equation}
Using the above definition, the Weinberg angle $\thetaW$ is given by
\begin{equation}
  \cos^2\thetaW = \frac{\muW^2}{\muZ^2}
\end{equation}
and the EW coupling reads
\begin{equation}
  \alphaEW = \frac{\sqrt{2}}{\pi} G_\mu \L|\muW^2 \sin^2\thetaW\R|.
\end{equation}
The phenomenological results of this paper were obtained setting the
Fermi constant to
\begin{equation}
  G_\mu = 1.1663787 \times 10^{-5}~\GeV^{-2}
\end{equation}
and the on-shell masses and widths of the $W$ and $Z$ bosons
to
\begin{equation}
  \begin{array}{l l l l l l}
    \mW^\OS = 80.379~\GeV & & & & & \GammaW^\OS = 2.085~\GeV
    \\ [1mm]
    \mZ^\OS = 91.1876~\GeV & & & & & \GammaZ^\OS = 2.4952~\GeV.
  \end{array}
\end{equation}
Following the prescription of ref.~\cite{Bardin:1988xt}, the
pole masses and widths of the $V = W, Z$ bosons are obtained from the
corresponding on-shell masses and widths as
\begin{equation}
  \label{pole_masses}
  \mV^2 = \frac{\L(\mV^\OS\R)^2}{\L(\mV^\OS\R)^2+\L(\GammaV^\OS\R)^2}
  \qquad
  \GammaV^2 =
  \frac{\L(\GammaV^\OS\R)^2}{\L(\mV^\OS\R)^2+\L(\GammaV^\OS\R)^2}.
\end{equation}
Finally, at order $\alphaS^2$ Feynman diagrams with Higgs boson
propagators appear. We set the mass $\mH$ and width $\GammaH$ of the
Higgs boson to
\begin{equation}
  \begin{array}{l l l l l l}
    \mH = 125~\GeV & & & & & \GammaH = 4.07 \times 10^{-3}~\GeV.
  \end{array}
\end{equation}

\subsection{Validation of fixed-order results}
\label{sec:validation}

\begin{table}[t]
\centering
\begin{tabular}{|c|c|c|c|c|}
\hline
$\sigma^{\text{NNLO}}_{q\bar{q}\to W^+W^-}$ [fb] & \Matrix & $\ptva=1~\mathrm{GeV}$ & $\ptva=5~\mathrm{GeV}$ & $\ptva=10~\mathrm{GeV}$ \\
\hline
$\mu=M_{4\ell}$ & $1328.0$ & $1329.2 \pm 8.1$ & $1331.8 \pm 5.7$ & $1330.7 \pm 4.2$\\
\hline
$\mu=M_{4\ell}/2$ & $1343.1$ & $1344.0 \pm 9.7$ & $1346.9 \pm 7.0$ & $1346.4 \pm 5.1$\\
\hline
$\mu=2M_{4\ell}$ & $1315.8$ & $1317.1 \pm 6.8$ & $1319.5 \pm 4.8$ & $1318.1 \pm 3.5$\\
\hline
\end{tabular}
\caption{Comparison of the \geneva and \Matrix results for the $q\bar{q}\to W^+W^-$ inclusive cross section. Results for different values of $\ptva$ are shown; we have set $n_f=4$ and used the \texttt{NNPDF31\_nnlo\_as\_0118\_nf\_4} PDF set~\cite{NNPDF:2017mvq}.}
\label{tab:nnlo_xs_validation}
\end{table}

We begin by validating the NNLO accuracy of our results for inclusive quantities. In \tab{nnlo_xs_validation}, we compare our NNLO predictions for the total cross section $pp\to W^+W^- \to e^- \bar{\nu}_e \mu^+ \nu_\mu$ at various values of $\ptva$ with results from the fixed order code \Matrix~\cite{Grazzini:2016ctr,Grazzini:2017mhc}. We note that both codes employ what is in essence a slicing method to reach NNLO accuracy -- in the case of the \geneva predictions, this is a slicing in $\ptja$ while in the \Matrix case the relevant variable is the transverse momentum of the colour singlet system, $q_T$. We therefore expect the results to differ from those of a fully local subtraction by amounts that are dependent on the size of the cut, due to missing $\mathcal{O}(\alpha_s^2)$ nonsingular contributions below the cut. Since \Matrix employs an extrapolation technique to $q_T^{\cut}=0~\GeV$ in order to minimise this effect, we assume that we are able to neglect any terms that may be missing at deeper orders in the power expansion, and use the difference between \geneva and \Matrix results to gauge the size of nonsingular power corrections in our formalism.

We note that the \geneva results are largely independent of the size of $\ptva$ up to $10~\GeV$ and show good agreement with the NNLO \Matrix predictions, both for central scale choices as well as when the scale is varied by a factor of two. This gives us confidence that, at our chosen value of $\ptva=1~\GeV$, we are able to safely neglect any power corrections. This removes the need for reweighting the below-cut contribution, a procedure that was followed in ref.~\cite{Alioli:2015toa} and which aimed to recover the integral of the missing terms. We also stress that the runs for each value of $\ptva$ were performed using the same number of CPU hours; the increase in statistical uncertainty with decreasing $\ptva$ is largely a consequence of the increased real-virtual matrix element evaluation time close to the singular limit.

\begin{figure}[tp]
\begin{center}
\begin{tabular}{cc}
\includegraphics[width=\rescaletwoplots]{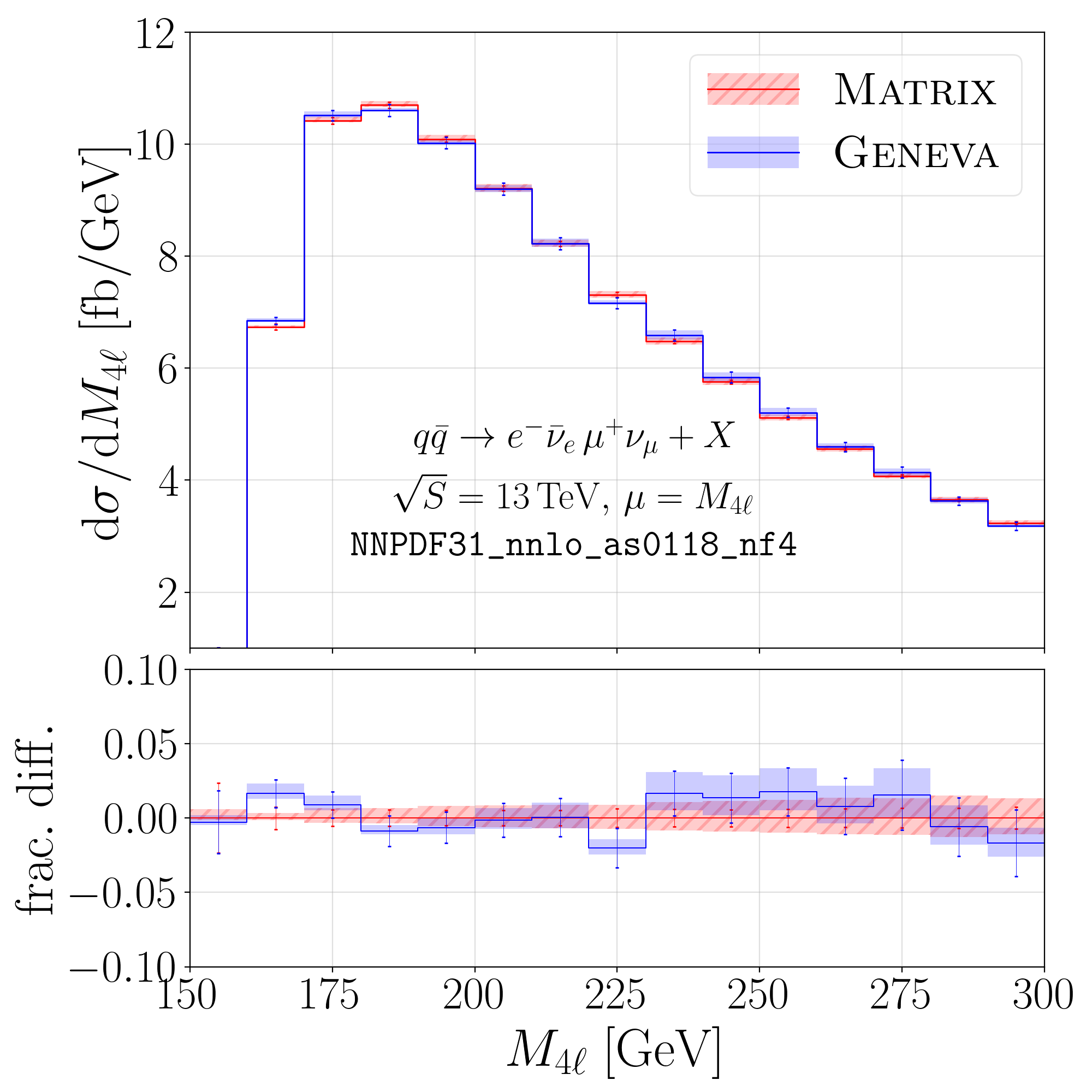} &
\includegraphics[width=\rescaletwoplots]{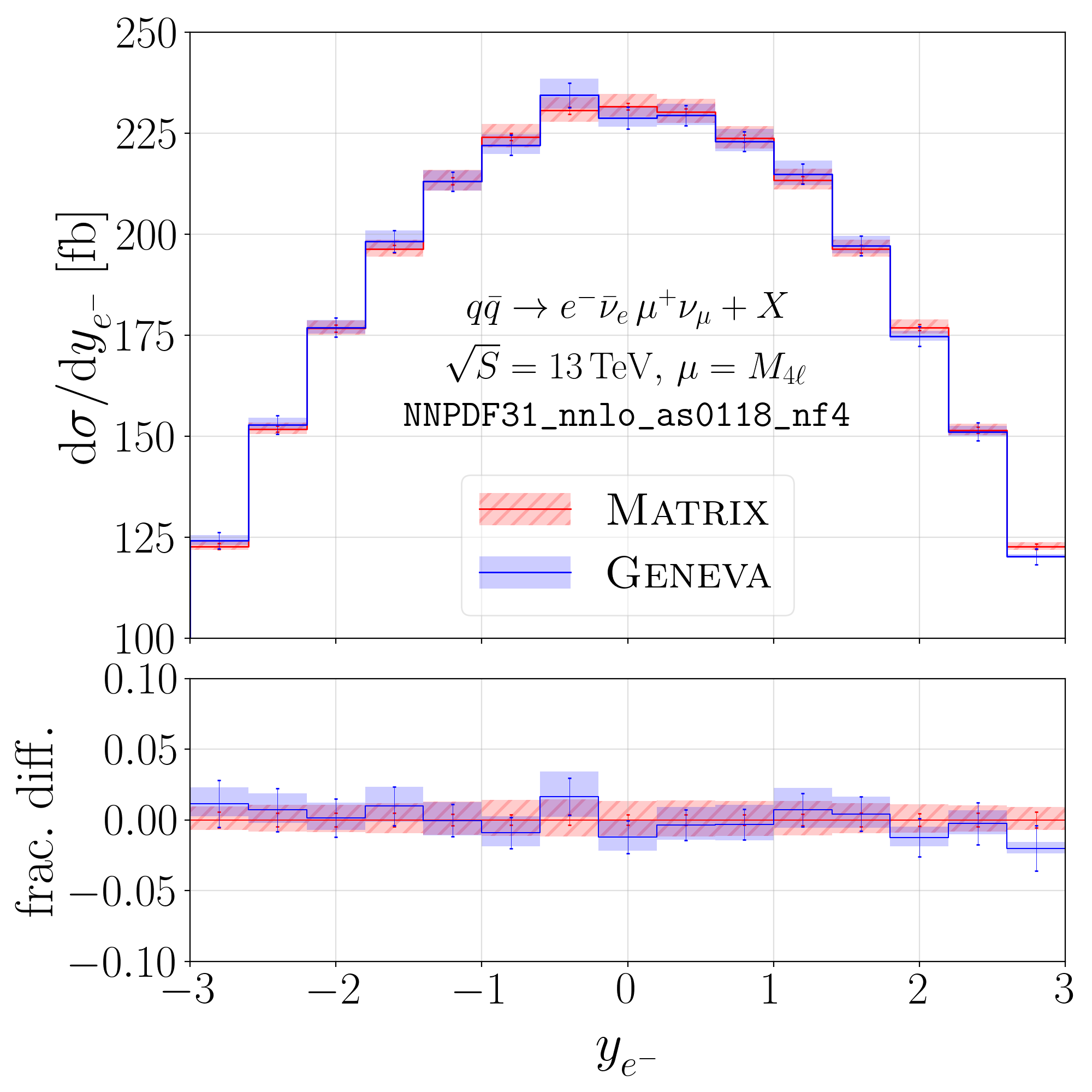} \\[\vspacebetweentwoplots]
\includegraphics[width=\rescaletwoplots]{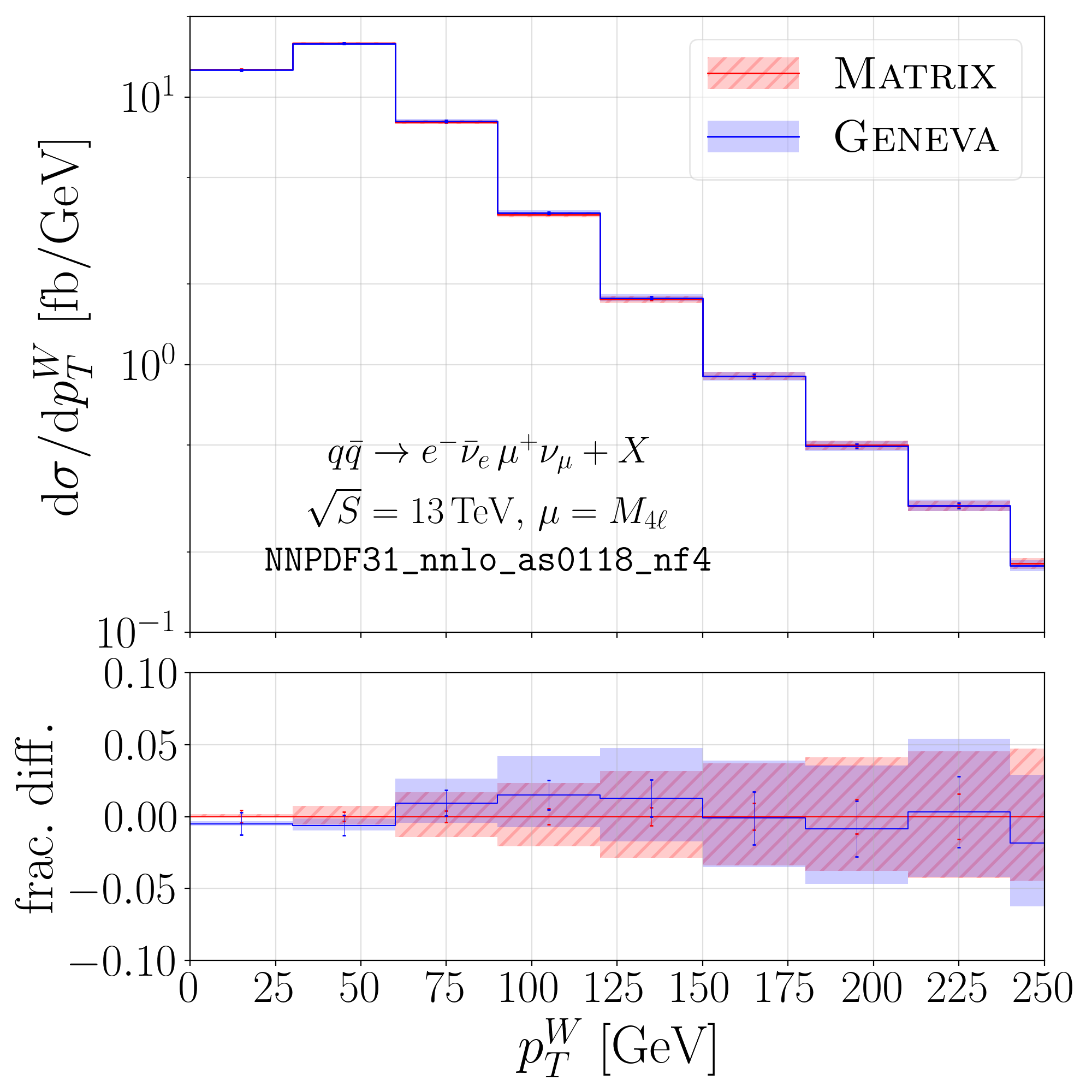} &
\includegraphics[width=\rescaletwoplots]{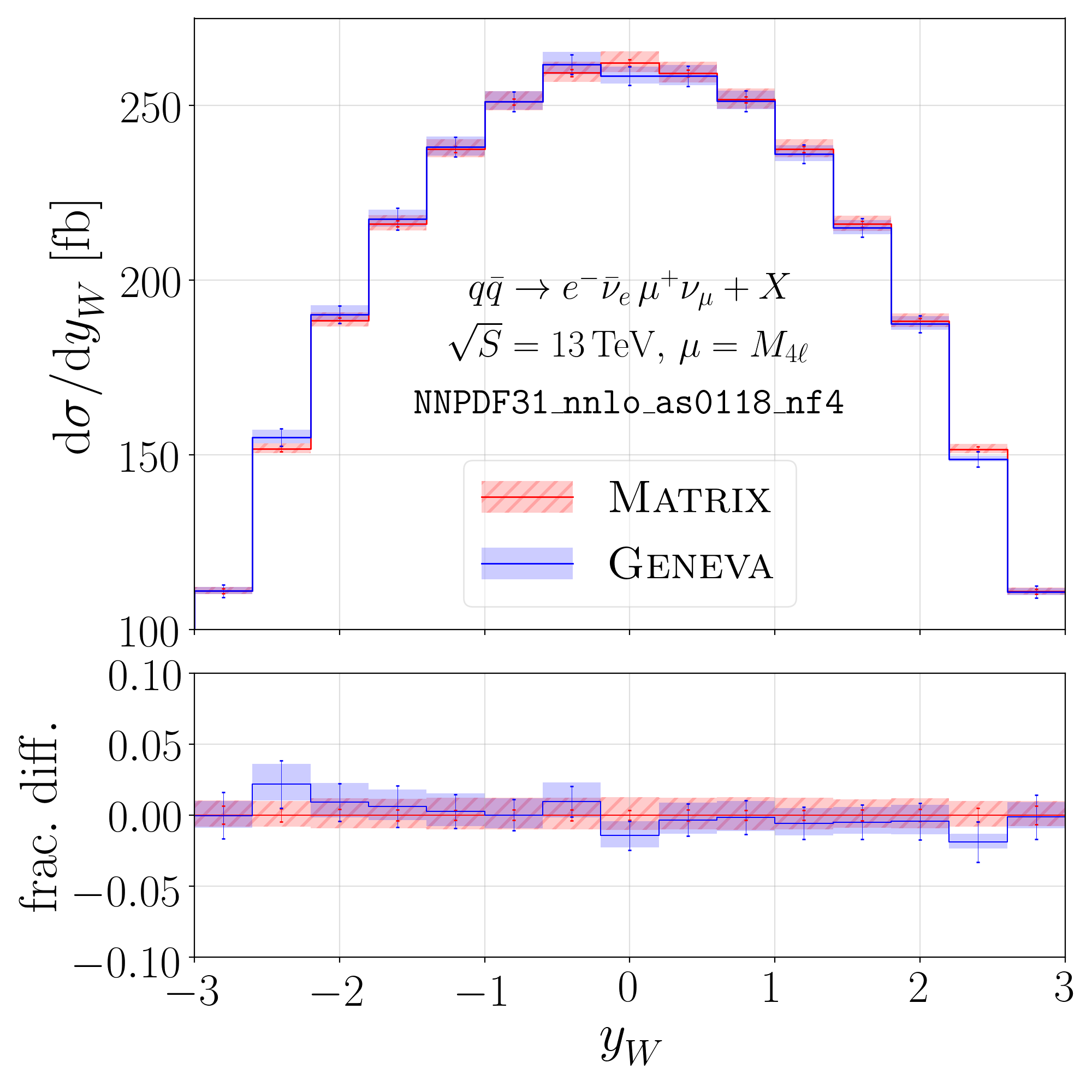} \\
\end{tabular}
\end{center}
\spaceabovefigurecaption
\caption{Comparison between \Matrix and \geneva for different kinematic distributions. We show the invariant mass of the colour-singlet system (top left), rapidity of the electron (top right), transverse momentum of the $W$ boson (bottom left) and rapidity of the $W$ boson (bottom right). Results have been obtained using the \texttt{NNPDF31\_nnlo\_as\_0118\_nf\_4} PDF set.
\label{fig:gvavsmatrixqqbar}
}
\spacebelowfigurecaption
\end{figure}

Turning to differential distributions, in \fig{gvavsmatrixqqbar} we show predictions for the invariant mass of the colour singlet and rapidity of the electron (top row) and for the transverse momentum and rapidity of the $W$ boson (bottom row) from \Matrix and \geneva. The \geneva results have been obtained setting $\ptva=\ptvb=1~\GeV$. We see good agreement for these inclusive distributions between the two calculations, thus validating the differential NNLO accuracy of our predictions. 

\subsection{Comparison to jet veto resummation in MCFM}
\label{sec:MCFM_validation}

\begin{figure}[t]
   \centering
   \includegraphics[width=\rescaletwoplots]{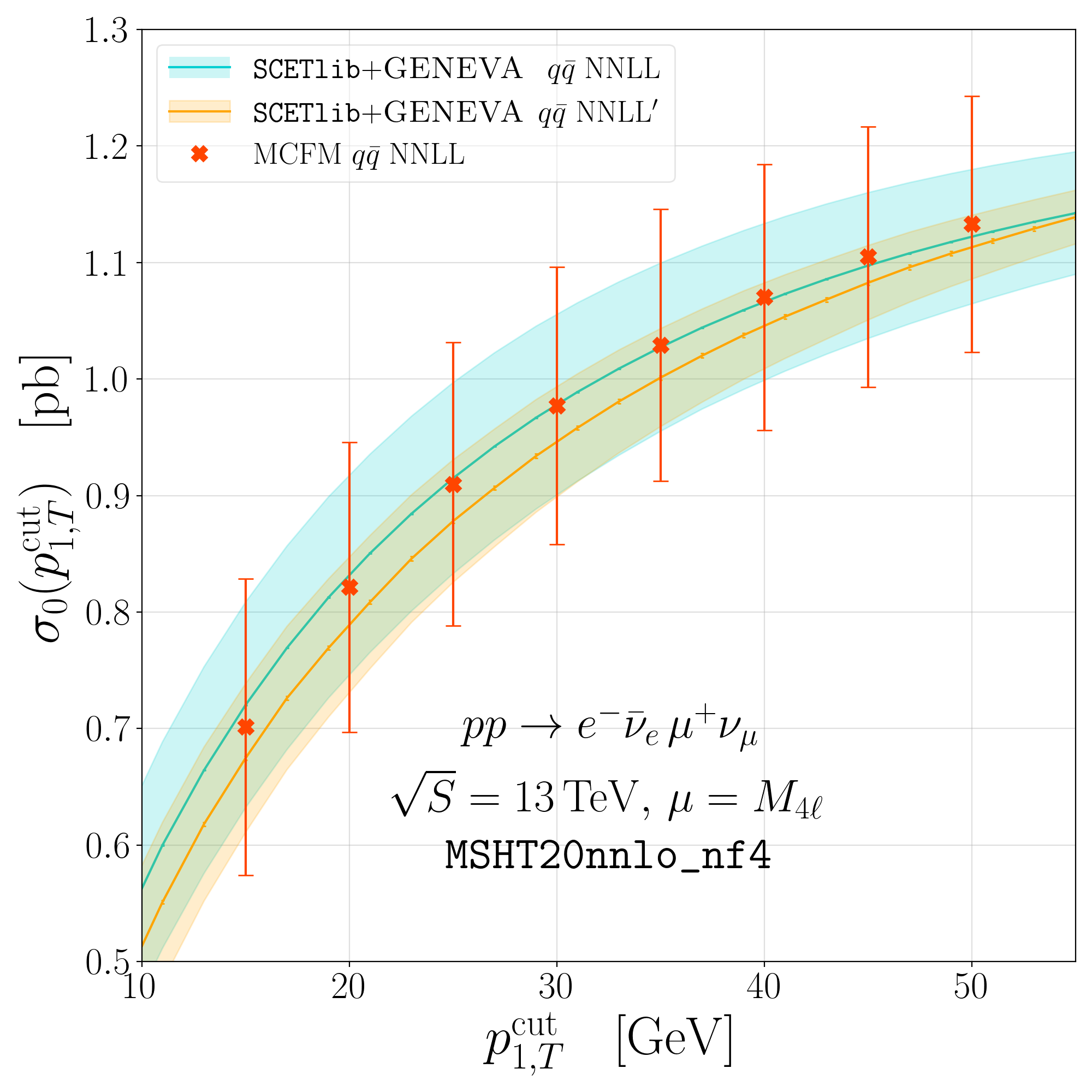}%
   \includegraphics[width=\rescaletwoplots]{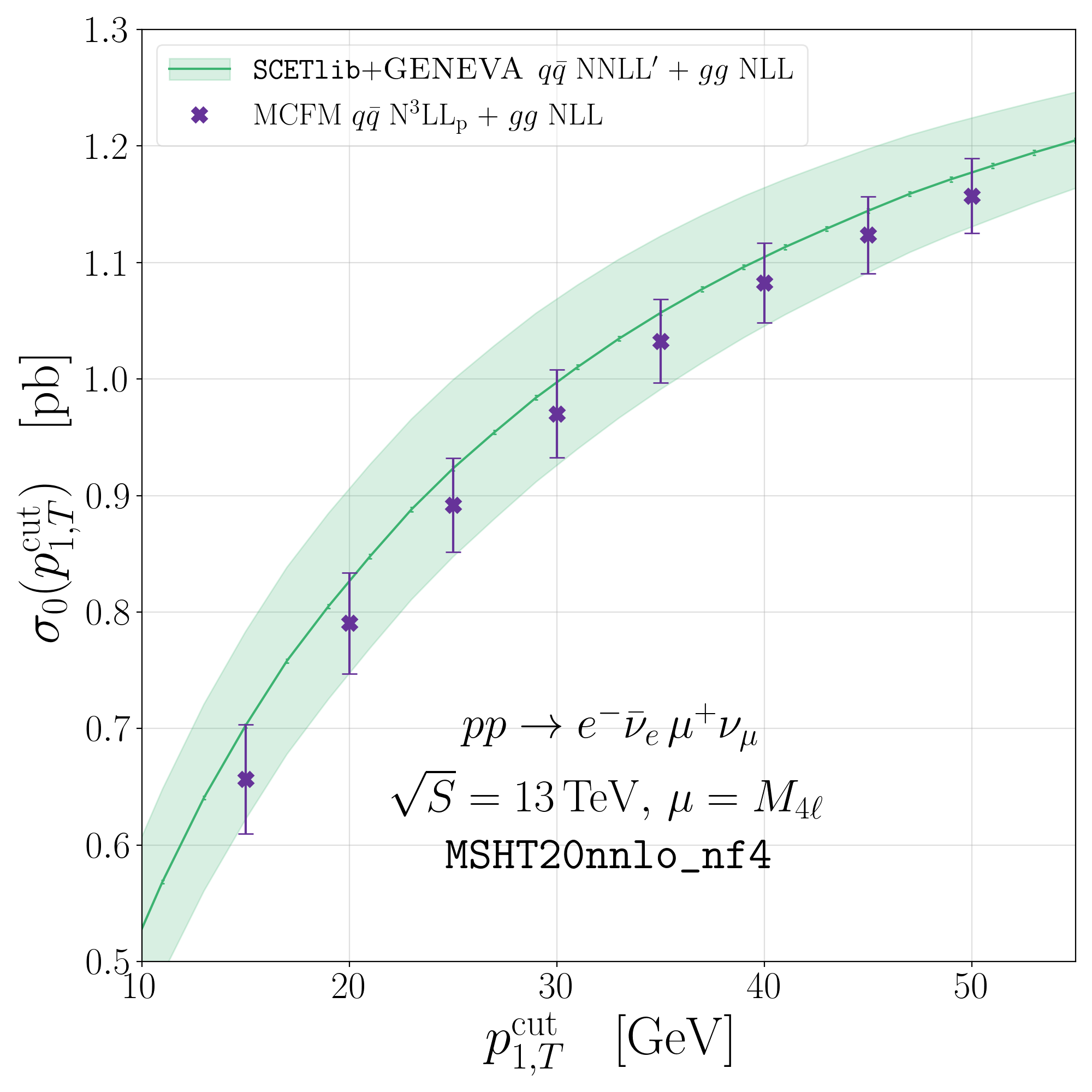}%
   \caption{Comparison of resummed results for the exclusive $0-$jet cross section from MCFM and \scetlib+\textsc{Geneva}. Successive orders of \geneva result are compared to the NNLL resummation from MCFM (left), and best predictions from both programs are also shown (right).}
   \label{fig:cmpMCFM}
\end{figure}
We have validated the logarithmic accuracy of our resummed predictions by comparison with MCFM version 10.3~\cite{Campbell:2023cha}. The program also adopts a SCET framework, implementing \eq{factorisation} to achieve up to approximate N$^3$LL (or N$^3$LL$_{\rm p}$) accuracy, but uses the notations and conventions of refs.~\cite{Becher:2012qa,Becher:2013xia,Becher:2014aya}. The primary difference with respect to the work in ref.~\cite{Stewart:2013faa}, upon which our own resummation is based, is the use of the collinear anomaly formalism to deal with rapidity divergences rather than the rapidity renormalisation group. We expect, however, the results from the two codes to be compatible at a given logarithmic order within theoretical uncertainties. 

In \fig{cmpMCFM} we compare the predictions for the exclusive $0-$jet cross section in $W^+W^-$ production from MCFM and \scetlib+~\geneva. We have calculated the MCFM uncertainties following the procedure described in ref.~\cite{Campbell:2023cha}, but for the purposes of comparison have omitted the variations of $R_0$ at N$^3$LL$_{\rm p}$. In the left panel, we show the \geneva results at NNLL and NNLL$'$ for the quark-initiated process, as well as MCFM results at NNLL. The \geneva results show a good convergence as the order is increased, with the NNLL$'$ band contained almost completely within that of the NNLL. We also see good agreement between the central \geneva predictions and those from MCFM, with small differences due to differing SCET conventions and our use of profile scales to switch off the resummation at larger $\ptva$. We note that the uncertainties for the MCFM predictions are considerably larger than ours at NNLL -- these are in fact dominated by the rapidity scale variations, which are varied up and down by a factor of 6~\cite{Campbell:2023cha}.

The right panel shows the best prediction from both codes for this observable, with \scetlib+~\geneva results at NNLL$'$ and MCFM at N$^3$LL$_{\rm p}$. In both cases, the gluon-initiated channel (which is loop-induced and begins to contribute at NNLO) has been included at NLL. Although one would not expect the two results to agree exactly, we observe consistent behaviour between the two calculations, with larger uncertainties in the \scetlib+~\geneva predictions and all MCFM points lying within the band. 

\subsection{Effect of NLL$'_{r_1}$ resummation}

\begin{figure}[t]
   \centering
   \includegraphics[width=\rescaletwoplots]{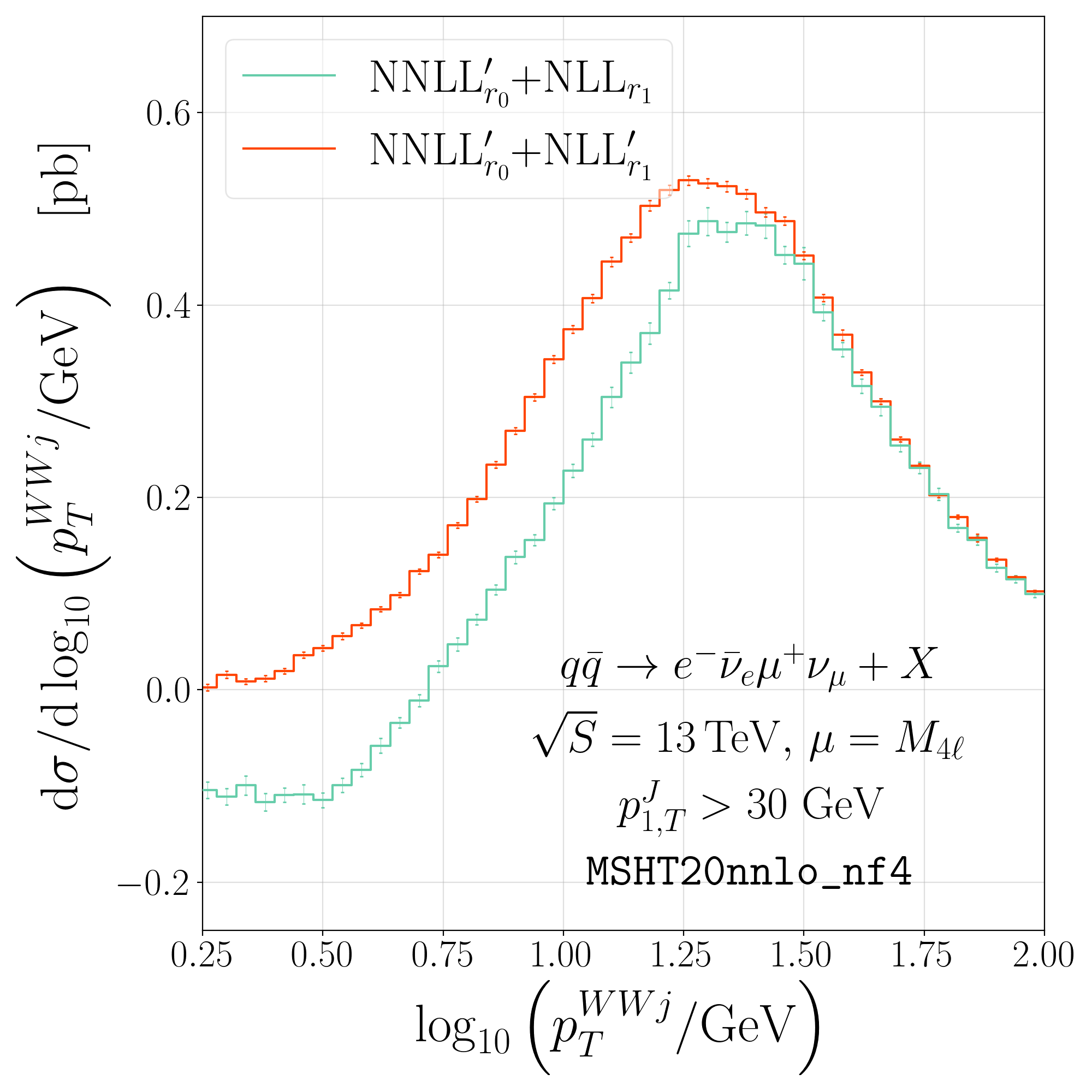}
   \caption{Effect of upgrading the resummation accuracy of the 1-jet resolution variable from NLL to NLL$'$. A cut $\ptja>30~\rm{GeV}$ has been placed on the first jet.}
   \label{fig:cmpprime}
\end{figure}

In \sec{NLLp1}, we described how the accuracy of the 1-jet resolution variable may be extended from NLL to NLL$'$ in the \geneva approach. \Fig{cmpprime} shows the effect of this extension on the transverse momentum of the system composed of the colour singlet and the hardest jet, when a minimum transverse momentum requirement is placed on the latter. We note that, compared to the NLL implementation, the NLL$'$ shows an improved cancellation at low values of $p_T^{WWj}$, indicating that the predictions for this quantity are now truly nonsingular in this limit. We have additionally verified that neither inclusive distributions nor the exclusive $0-$jet cross section $\sigma_0(\ptva)$ are altered by this change. All subsequent results shown in this work will make use of the NLL$'$ resummation for $\ptjb$.

\subsection{Showered results}
\label{sec:showeredresults}
\begin{figure}[ht]
   \centering
   \includegraphics[width=\rescaletwoplots]{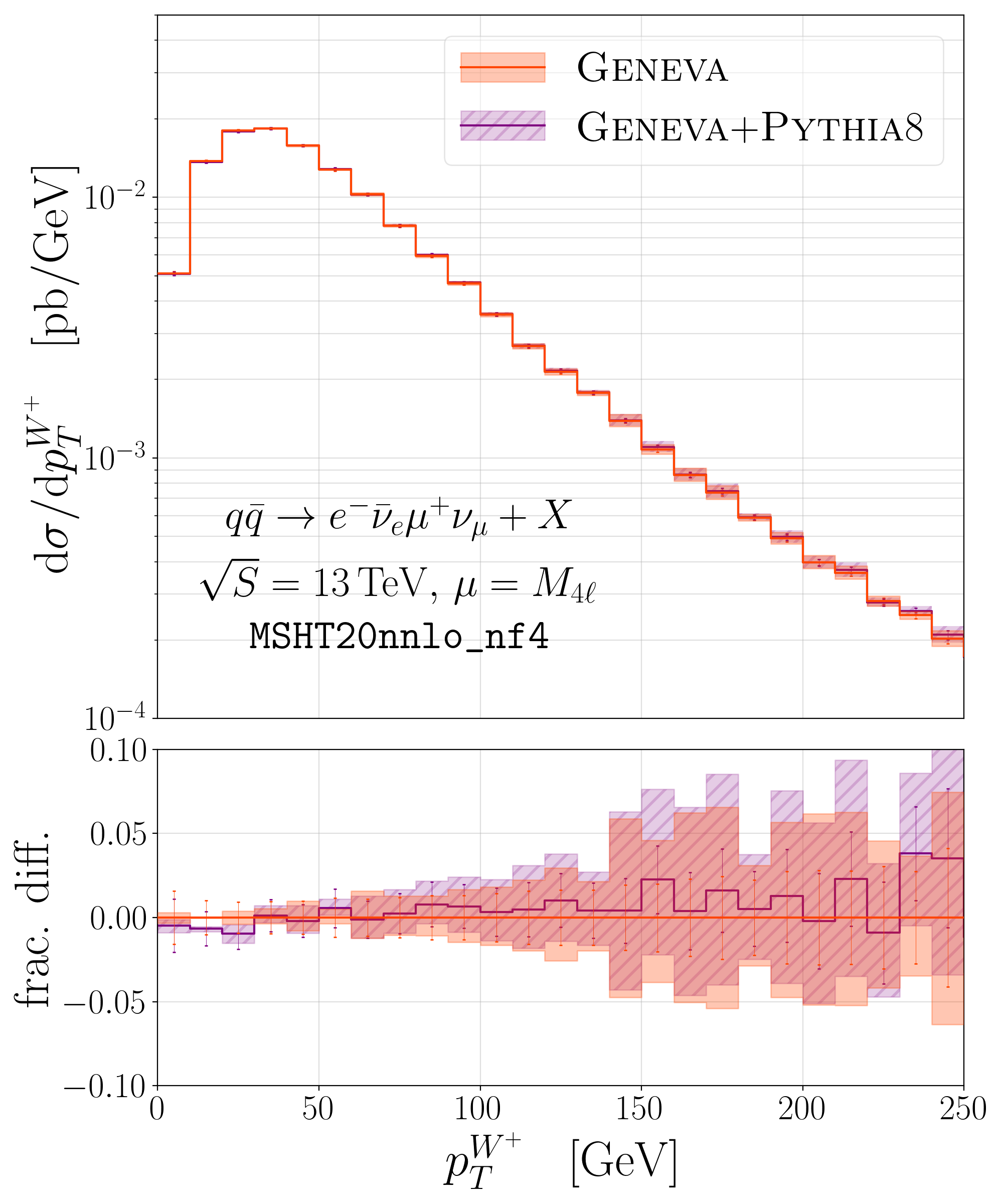}%
   \includegraphics[width=\rescaletwoplots]{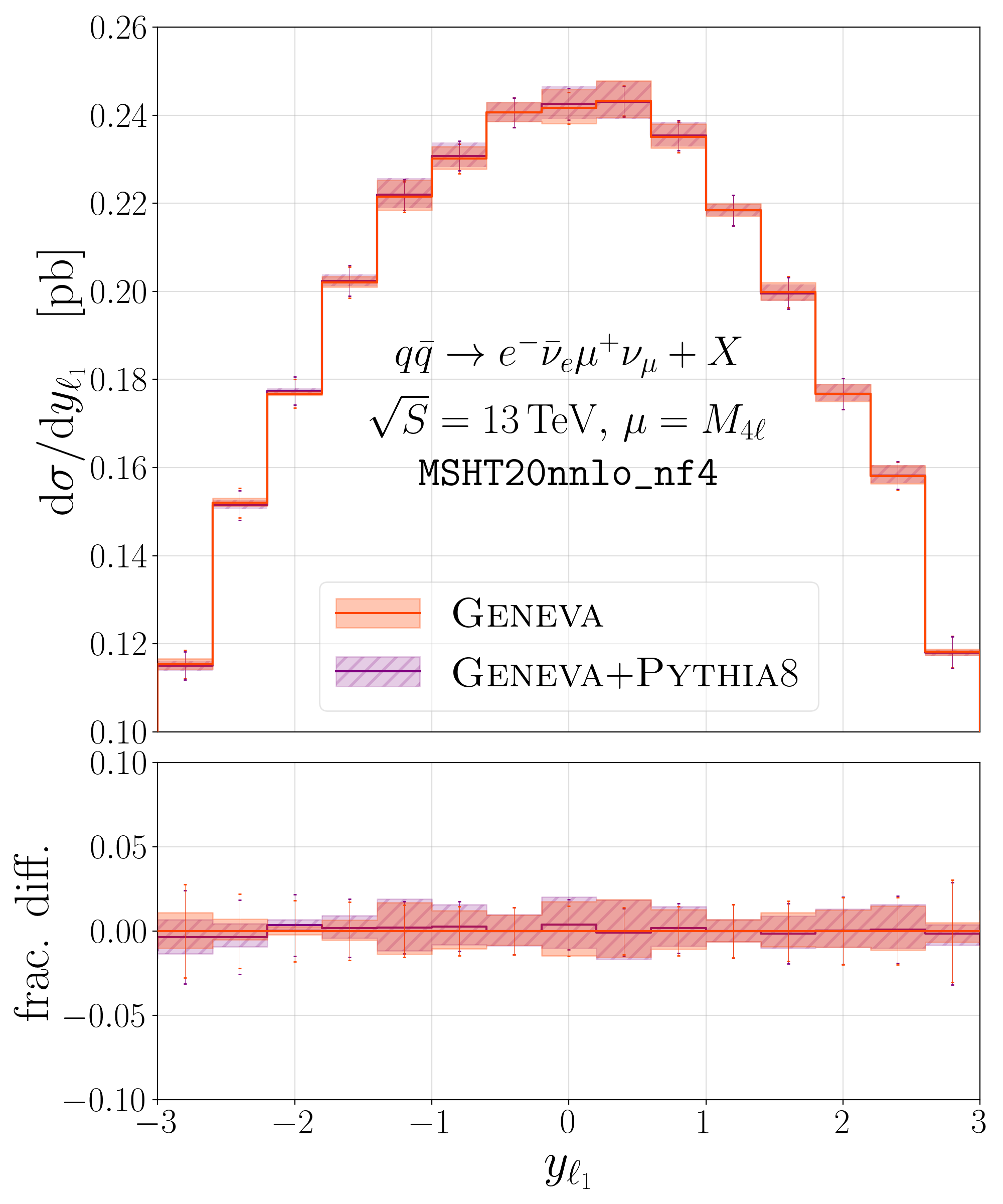}%
   \caption{Comparison of inclusive distributions, before (red) and after (purple) showering with \pythia. The transverse momentum of the $W^+$ boson (left) and the rapidity of the hardest charged lepton (right) are shown.}
   \label{fig:cmpinclusive}
\end{figure}

\begin{figure}[t]
   \centering
   \includegraphics[width=\rescaletwoplots]{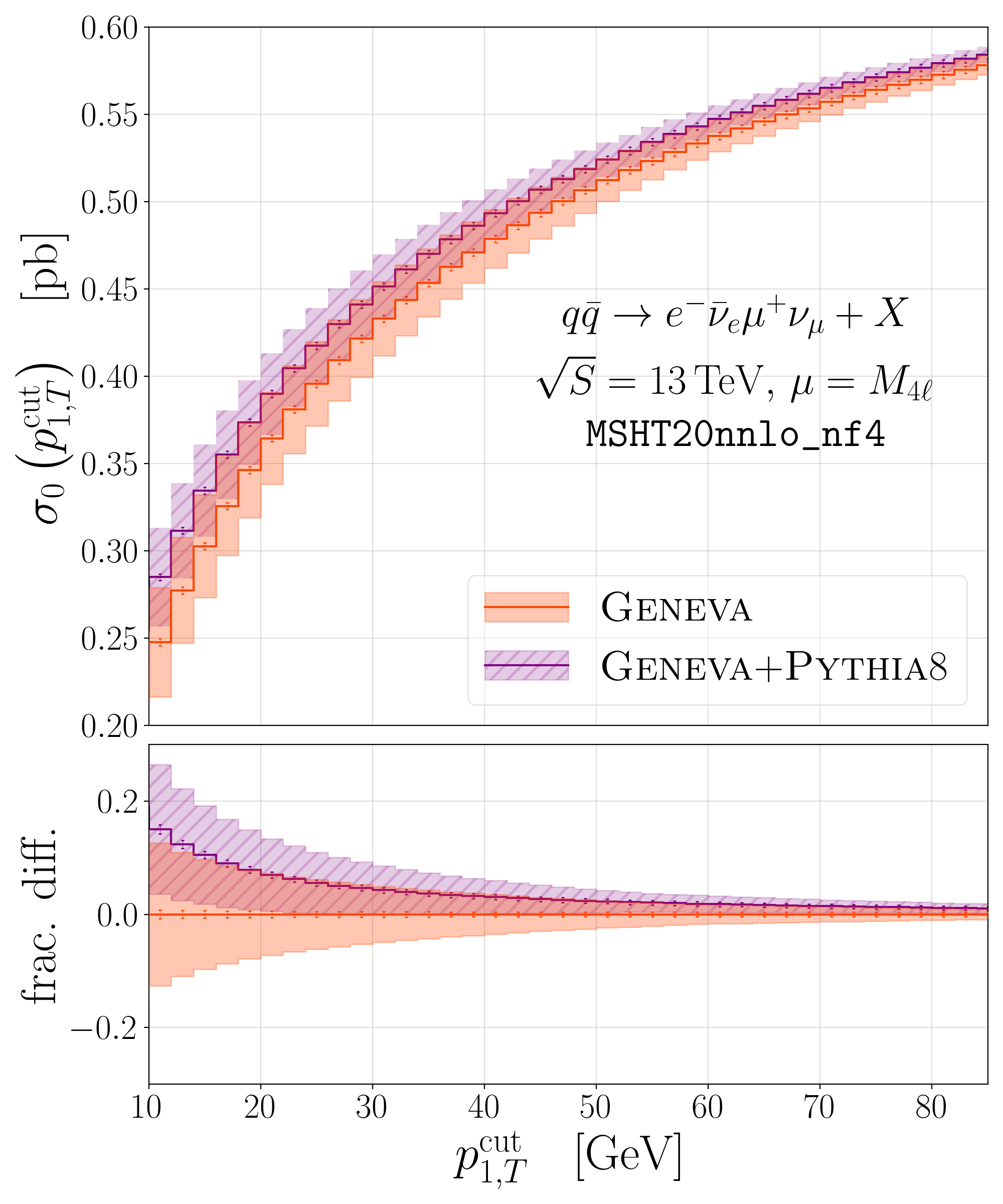}
   \caption{Comparison of the exclusive 0-jet cross section, before (red) and after (purple) showering with \pythia.}
   \label{fig:cmpexclusive}
\end{figure}

In \fig{cmpinclusive}, we present two examples of inclusive distributions, before and after showering with \textsc{Pythia8}~\cite{Sjostrand:2007gs}. For the purposes of this comparison, QED effects in the shower, hadronisation and multiple parton interactions have been deactivated. The left panel shows the transverse momentum of  the $W^+$ boson, while the right panel shows instead the rapidity of the hardest charged lepton. We observe that the shower does not affect these distributions, as expected, and there is good agreement for both the central values and the scale variation bands. In \fig{cmpexclusive}, we present the $0-$jet exclusive cross section as a function of $\ptva$, again showing the distribution before and after adding the effect of the parton shower. We do not necessarily expect the accuracy of this distribution to be preserved by the shower; however, we observe a reasonable agreement between the curves which is within the scale uncertainty bands, and which improves with increasing $\ptva$. In particular, for the values of the jet veto typically imposed by experimental analyses $\sim 30~\GeV$, we notice an upwards shift in the cross section of $\sim 3\%$.

\subsection{Comparison to experimental data}
\label{sec:expdata}

\begin{table}[t!]
\centering
\begin{tabular}{c|c|c}
& ATLAS~\cite{ATLAS:2019rob} & CMS~\cite{CMS:2020mxy} \\
\hline
$p_T^{\ell}$ & $>27~\GeV$ & $>20~\GeV$ \\
$|\eta^{\ell}|$ & $<2.5$ & $<2.5$ \\
$m_{\ell\ell}$ & $> 55~\GeV$ & $> 20~\GeV$ \\
$p_T^{\ell\ell}$ & $>30~\GeV$ & $> 30~\GeV$ \\
$E_T^{\rm miss}$ & $>20~\GeV$ & $> 30~\GeV$ \\
No jets with $\{p_T, |\eta|\}$ & $\{>35~\GeV, <4.5\}$ & $\{>30~\GeV, <4.5\}$ \\
\end{tabular}
\caption{Definition of the fiducial phase space in each of the experimental analyses considered in this work.}
\label{tab:fiducial_ps}
\end{table}

We now present the comparison to the measurements of $W^+W^-$ production taken by the ATLAS~\cite{ATLAS:2019rob} and CMS~\cite{CMS:2020mxy} experiments at a centre-of-mass energy of 13 TeV, corresponding to integrated luminosities of 36.1 and 35.9 fb$^{-1}$ respectively. Both analyses employ a jet veto, with jets defined using the anti-$k_T$ algorithm~\cite{Cacciari:2008gp} with $R=0.4$; details of the fiducial regions designated in each case are provided in \tab{fiducial_ps}.\footnote{We note that the ATLAS collaboration has recently published a further measurement of the process in ref.~\cite{ATLAS:2023zis}; since no jet veto is employed in this analysis, however, we do not consider it here.} We generate events for the process $p p \to (W^+ \to \mu^+ \nu_\mu) (W^- \to e^- \bar{\nu}_e)$, i.e. the different-flavour channel. In the case of the ATLAS measurement, both this channel and the charge conjugated channel are considered -- we therefore rescale our results by a factor of two. The CMS measurement, however, considers also the same-flavour channels. Following ref.~\cite{Campbell:2023cha}, we multiply our results by a factor of 4.15 to account for the slight enhancement due to the contribution to the same-flavour process arising from $ZZ$ production. Furthermore, when comparing with both ATLAS and CMS we multiply the $gg$-initiated contribution by the inclusive NLO $k$-factor of 1.7 used in the ATLAS analysis for other event generators~\cite{Caola:2016trd}. In the absence of a full NLO \geneva implementation, one should in principle rescale by a $k$-factor determined in the correct fiducial region for each experiment. However, to allow a more direct comparison with the results of other generators presented in ref.~\cite{ATLAS:2019rob}, we follow their prescription here. When considering kinematic distributions, CMS presents only normalised measurements -- for consistency, we therefore consider only normalised ATLAS data. We perform our comparisons using the analyses as implemented in \textsc{Rivet}~\cite{Buckley:2010ar}, setting $\ptva=1~\GeV$ and $\ptvb=0.5~\GeV$.

\begin{figure}[ht!]
   \centering
   \includegraphics[width=\rescaletwoplots]{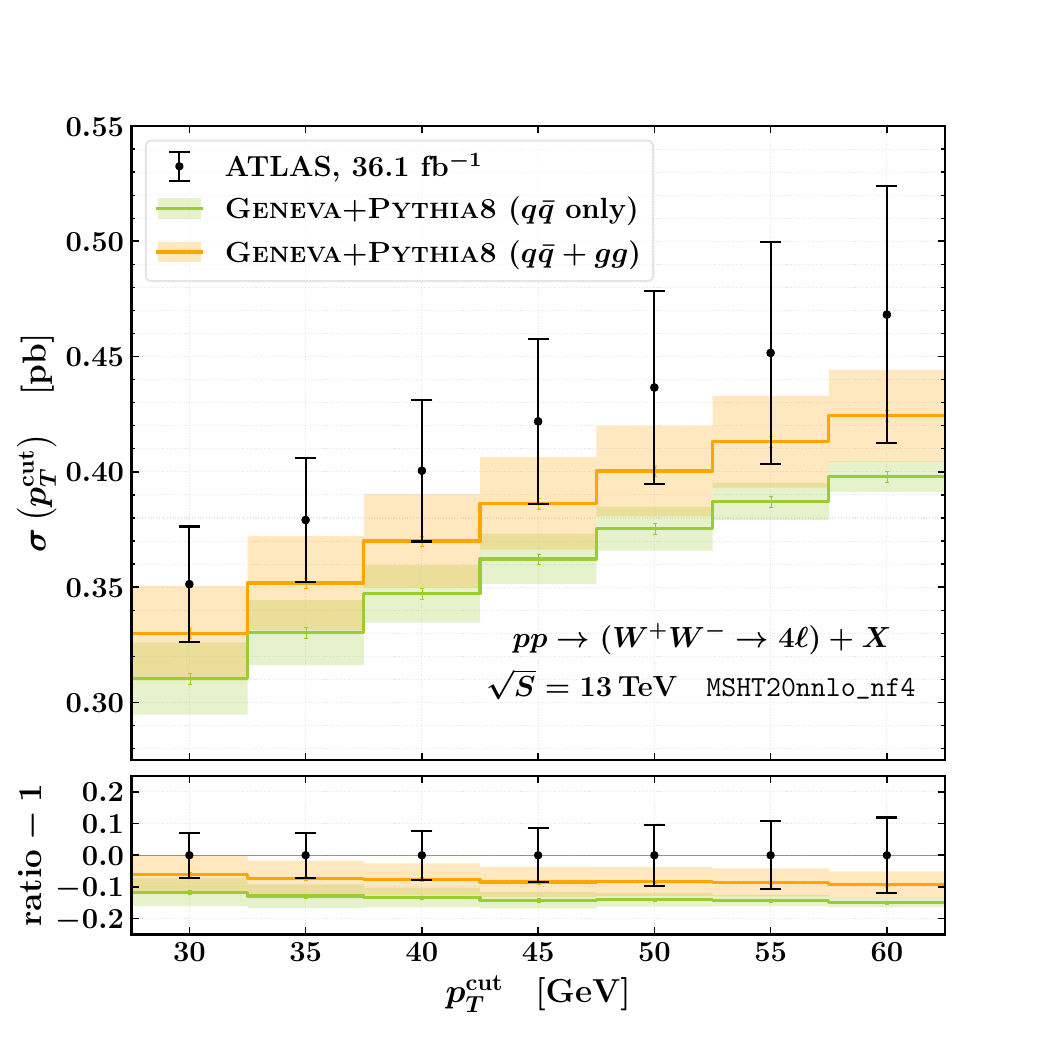}%
   \includegraphics[width=\rescaletwoplots]{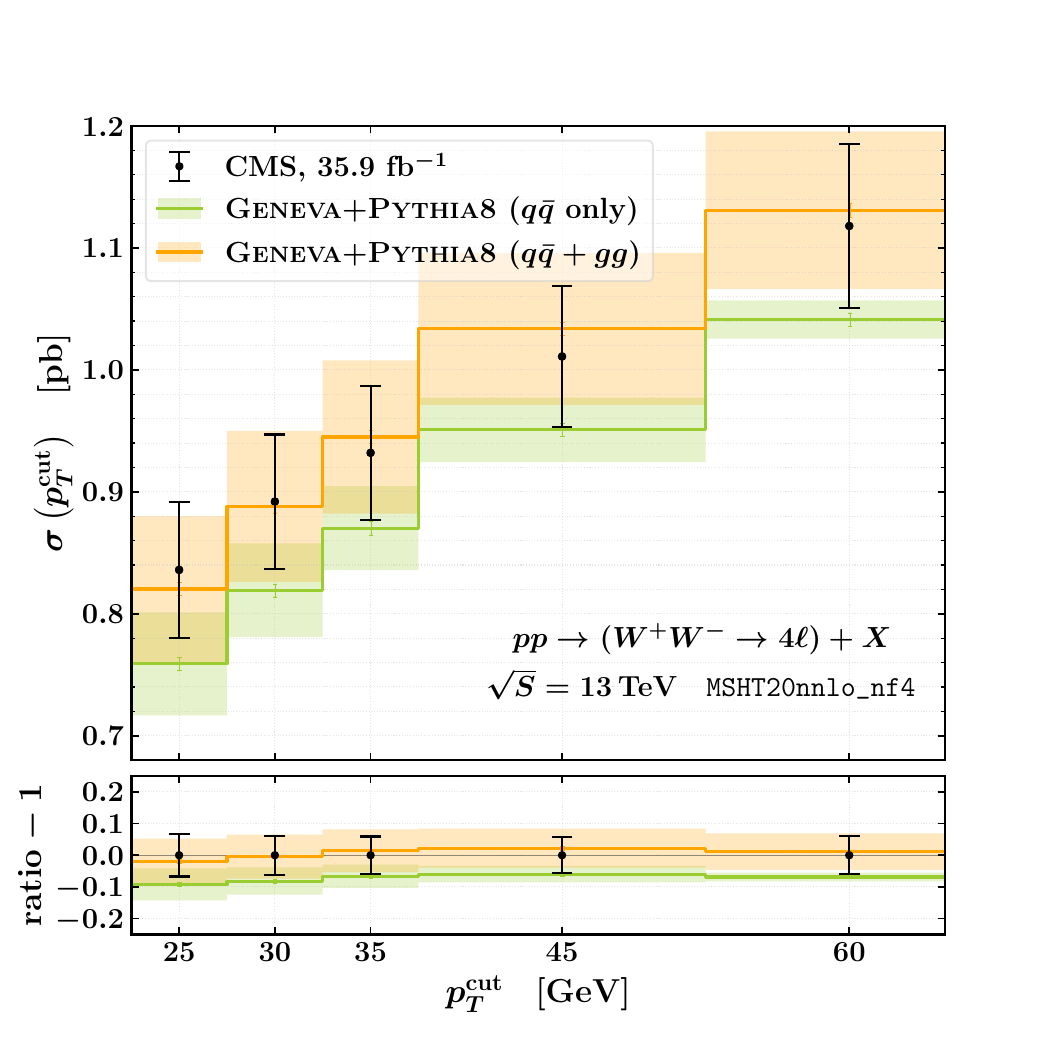}%
   \caption{Comparison of \geneva predictions for the exclusive $0-$jet cross section as a function of $\ptv$, against ATLAS (left) and CMS (right) data taken at 13 TeV. Results for the $q\bar{q}$-initiated channel are shown alone, as well as in combination with the $gg$-initiated channel.}
   \label{fig:ATLASCMSsigmazerojet}
\end{figure}

In \fig{ATLASCMSsigmazerojet}, we compare our predictions for the exclusive $0-$jet cross section against ATLAS and CMS measurements. We show predictions for the $q\bar{q}$ channel alone, as well as in combination with the gluon-initiated channel. We note first that the inclusion of the $gg$ channel is necessary in order to obtain a good description of the data. Since the channel opens only at $\mathcal{O}(\alpha_s^2)$ with respect to the $q\bar{q}$ channel, it is effectively a `leading order' contribution and the scale variations associated with it are correspondingly large. In the ATLAS case the overall normalisation seems to be incorrectly predicted, resulting in the \geneva predictions slightly undershooting the data for all values of the jet veto (though in all cases consistent within uncertainties). Similar behaviour was observed for other event generators in ref.~\cite{ATLAS:2019rob}. In the CMS case, however, we observe excellent agreement in both shape and normalisation.

\begin{figure}[tp]
\begin{center}
\begin{tabular}{cc}
  \includegraphics[width=\rescaletwoplots]{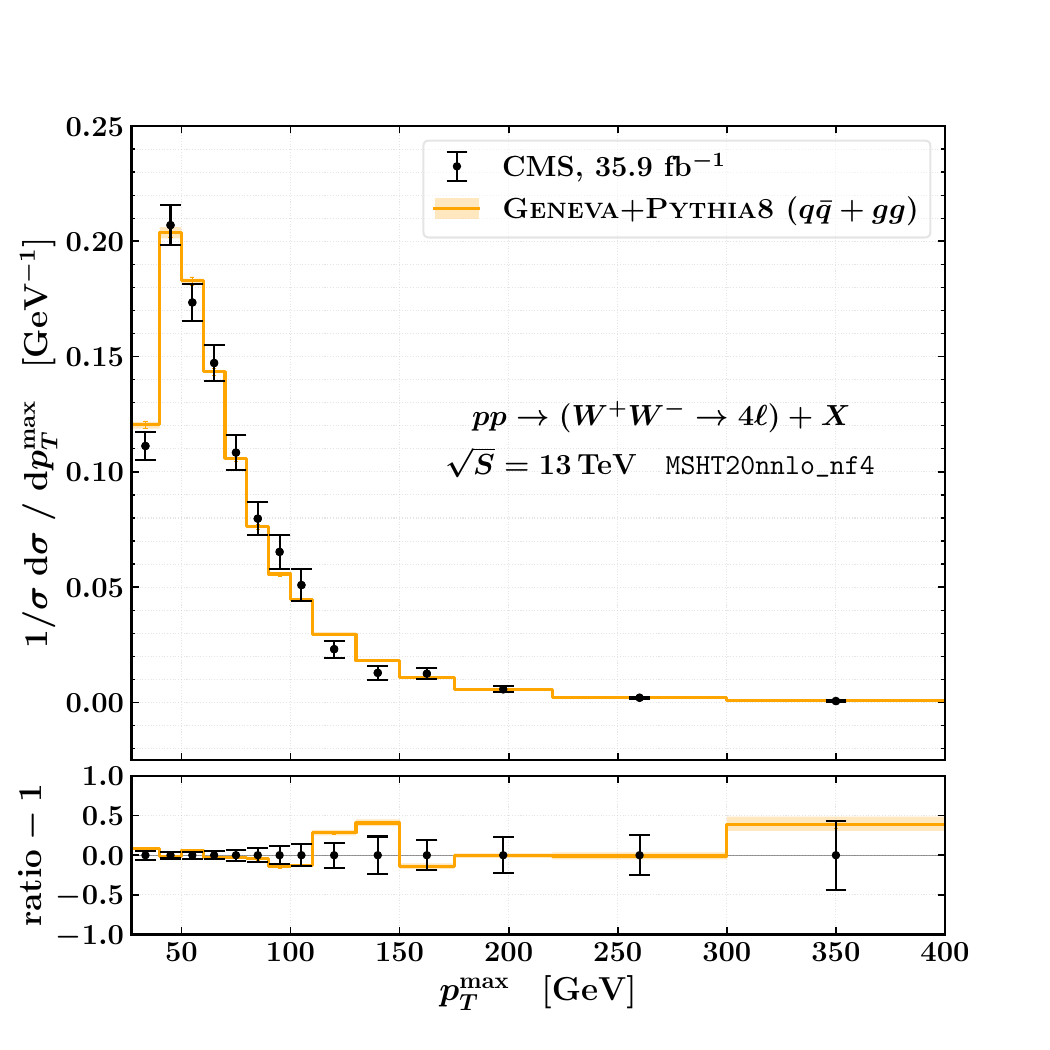} &
  \includegraphics[width=\rescaletwoplots]{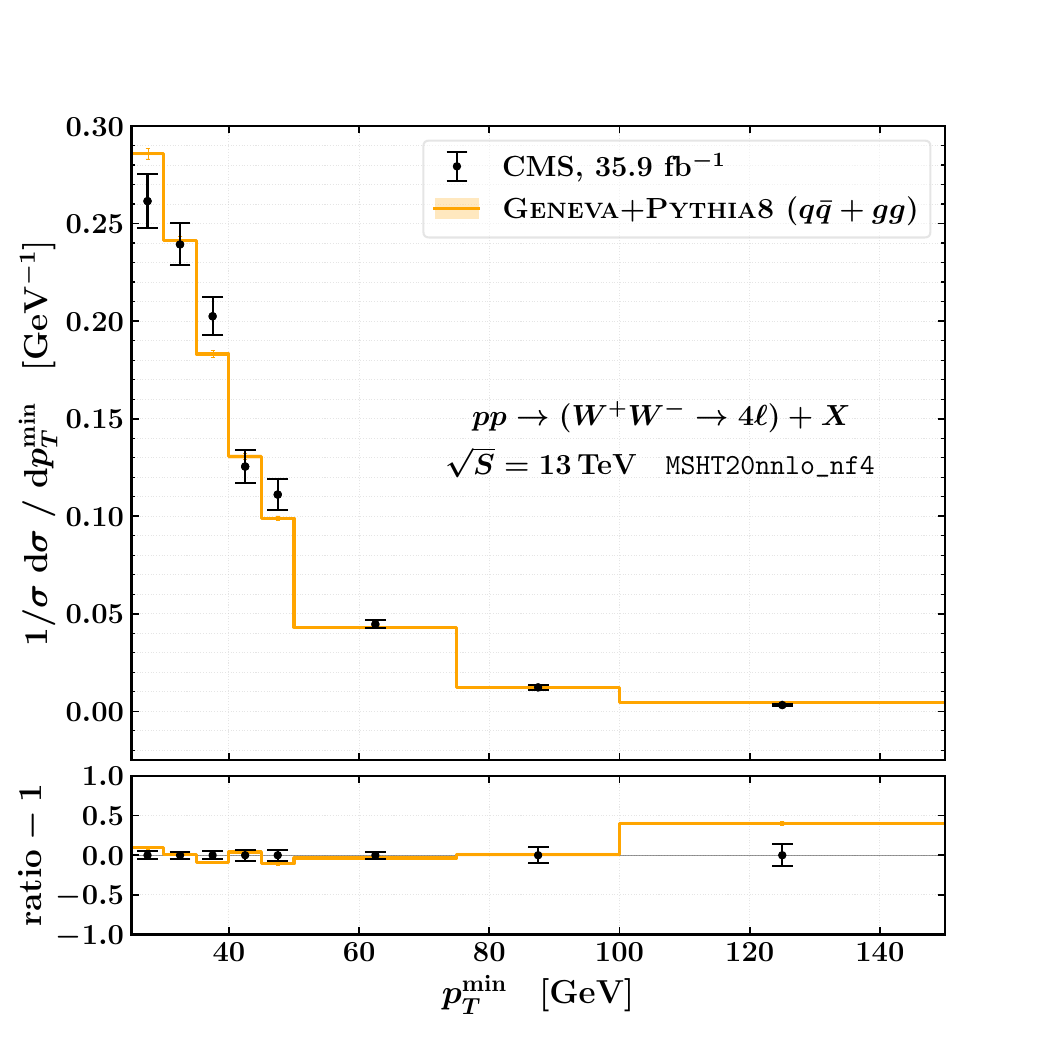}
  \\ \includegraphics[width=\rescaletwoplots]{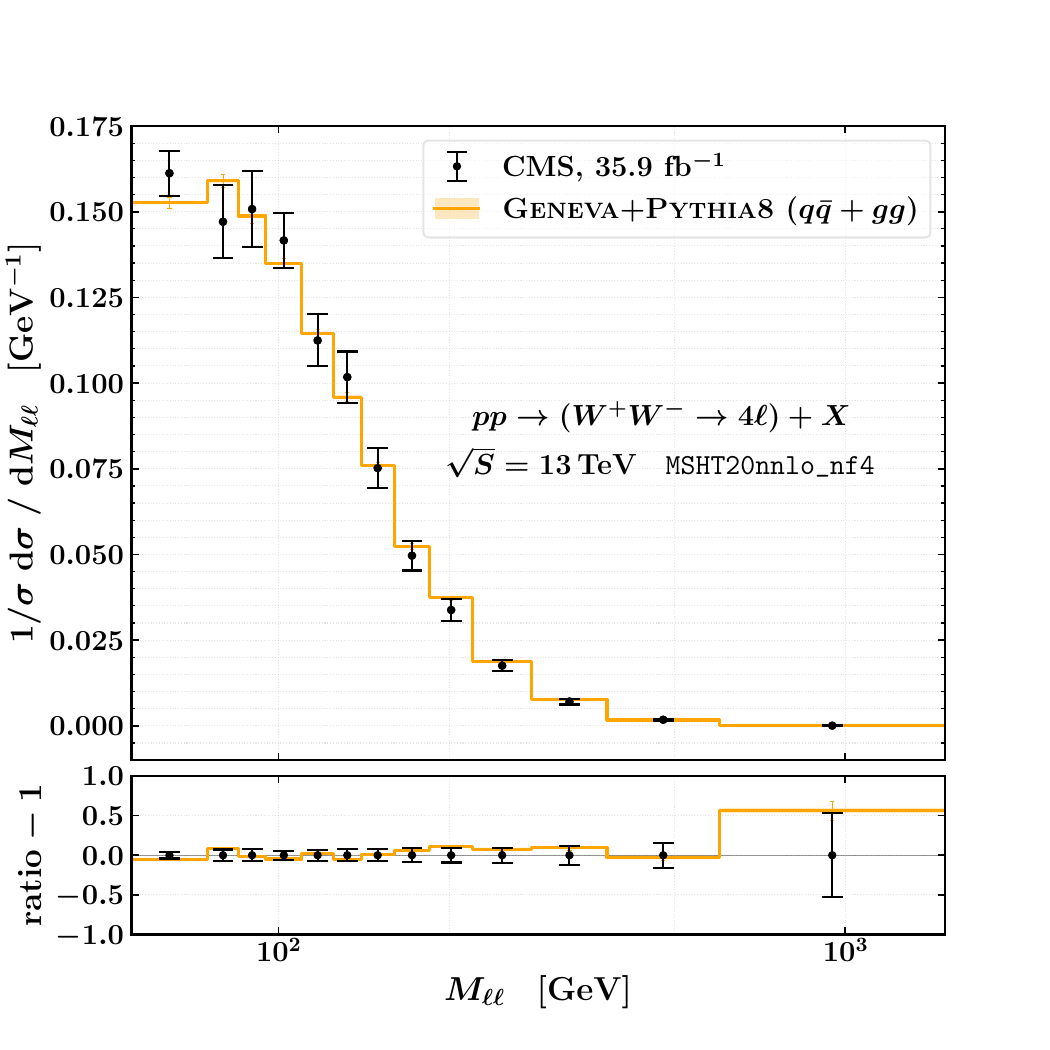} &
  \includegraphics[width=\rescaletwoplots]{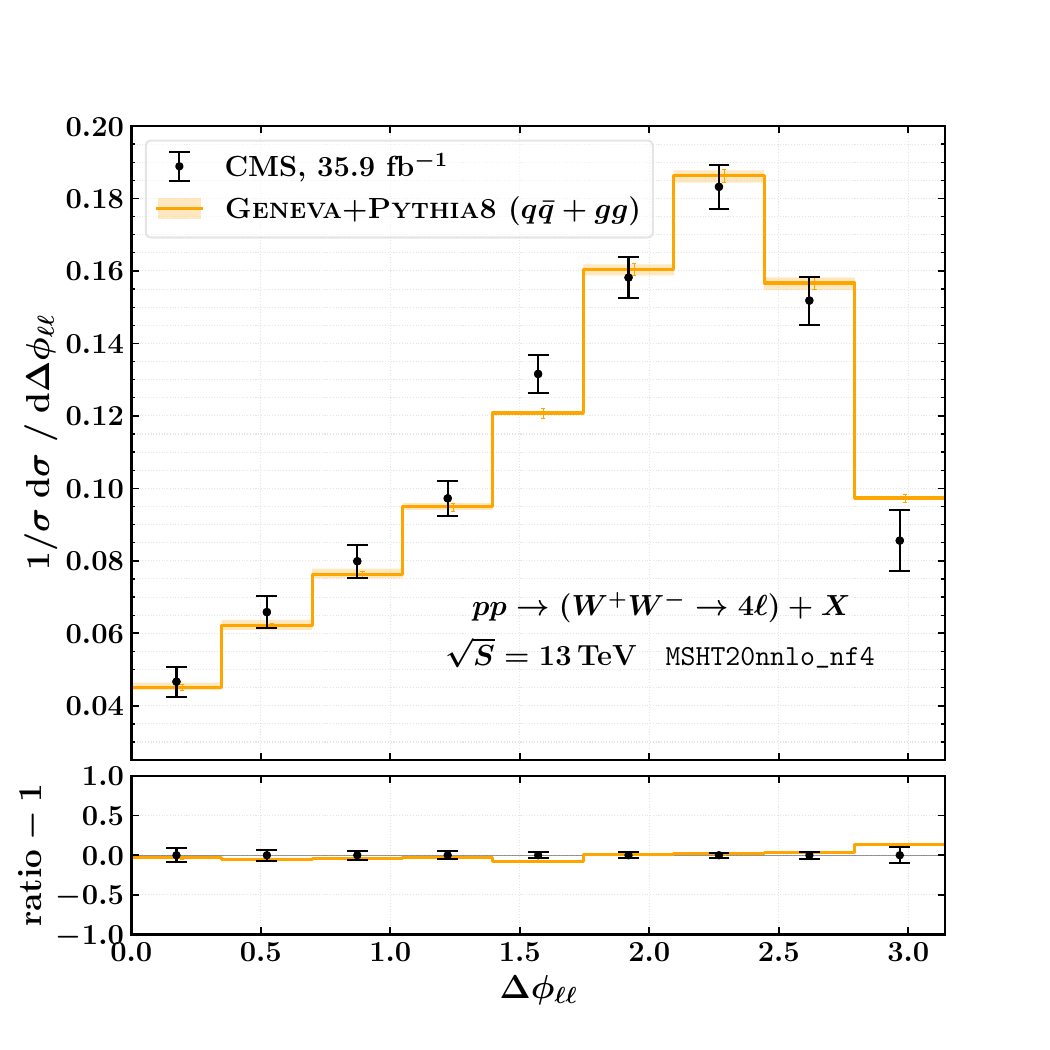} \\
\end{tabular}
\end{center}
\spaceabovefigurecaption
\caption{Comparison of \geneva predictions against CMS data for different kinematic distributions. We show the maximum and minimum lepton transverse momentum in the top left and right, and the invariant mass and azimuthal separation of the charged lepton pair in the bottom left and right.
\label{fig:CMSkinematic}
}
\spacebelowfigurecaption
\end{figure}

\begin{figure}[ht!]
   \centering
   \includegraphics[width=\rescaletwoplots]{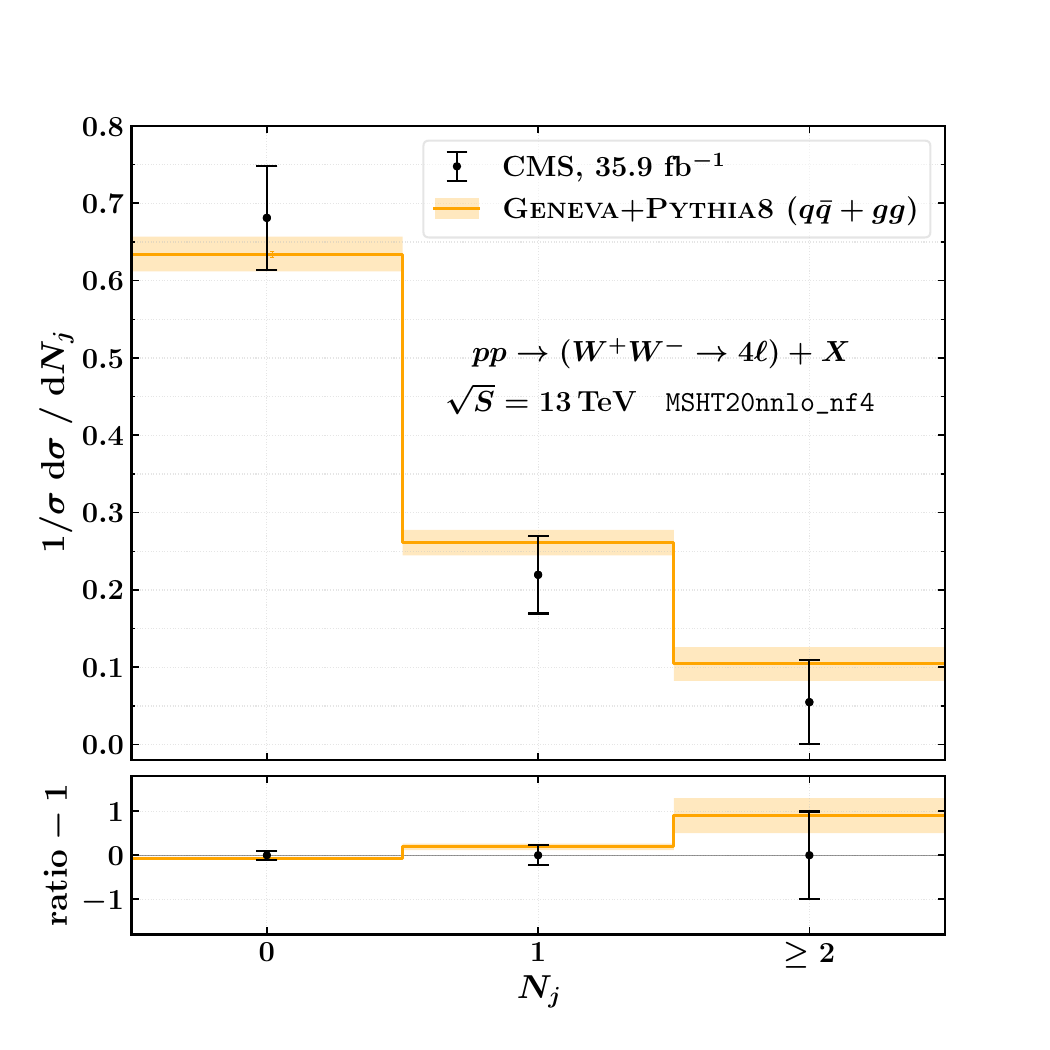}
   \caption{Comparison of \geneva predictions against CMS data for the jet multiplicity.
   \label{fig:CMSjetmult}
   }
\end{figure}

\Fig{CMSkinematic} shows instead four different normalised kinematic distributions as measured by CMS: the transverse momentum of the harder and softer charged lepton, the invariant mass of the charged lepton pair and the azimuthal separation of the charged leptons. We observe a good description of the data, with the exception of the high invariant mass/transverse momentum bins where electroweak corrections are likely to play an important r\^{o}le~\cite{Grazzini:2019jkl}. Predictions against data for the jet multiplicity measurement are shown in \fig{CMSjetmult}: we observe good agreement in all bins. In this case again the size of the scale uncertainty is driven by the $gg$-initiated channel, motivating the inclusion of genuine NLO corrections to this subprocess (rather than the simple $k$-factor rescaling we have performed here). 

\begin{figure}[tp]
\begin{center}
\begin{tabular}{cc}
\includegraphics[width=\rescaletwoplots]{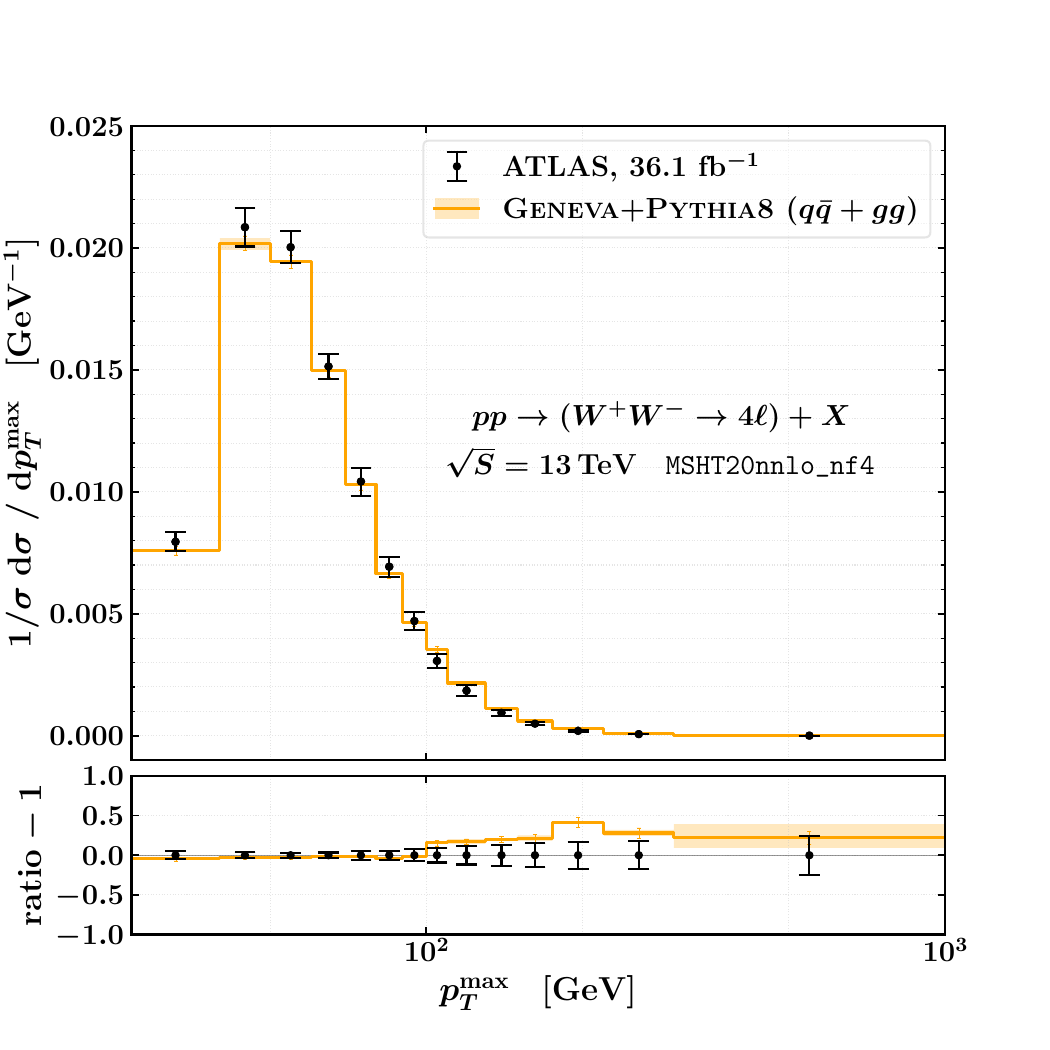} &
\includegraphics[width=\rescaletwoplots]{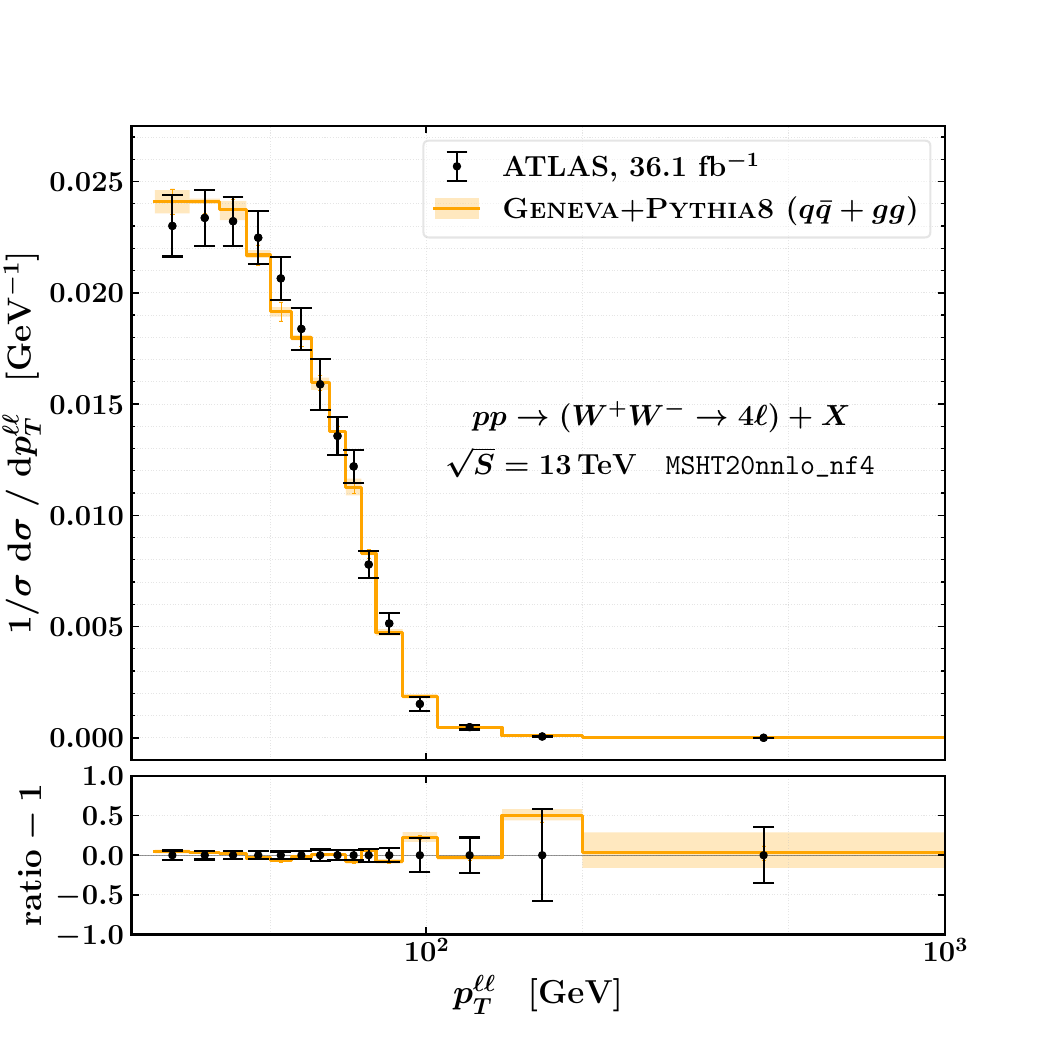}
\\ \includegraphics[width=\rescaletwoplots]{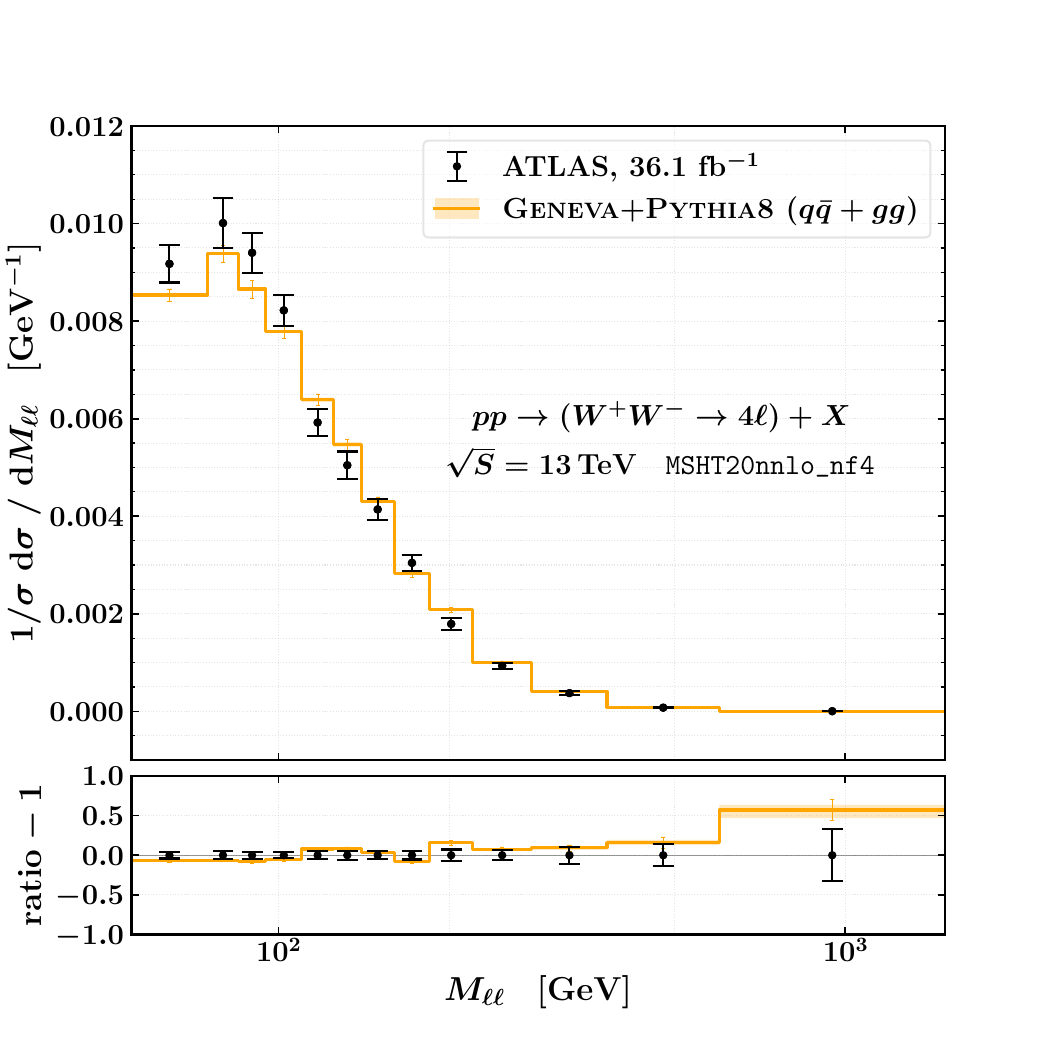} &
\includegraphics[width=\rescaletwoplots]{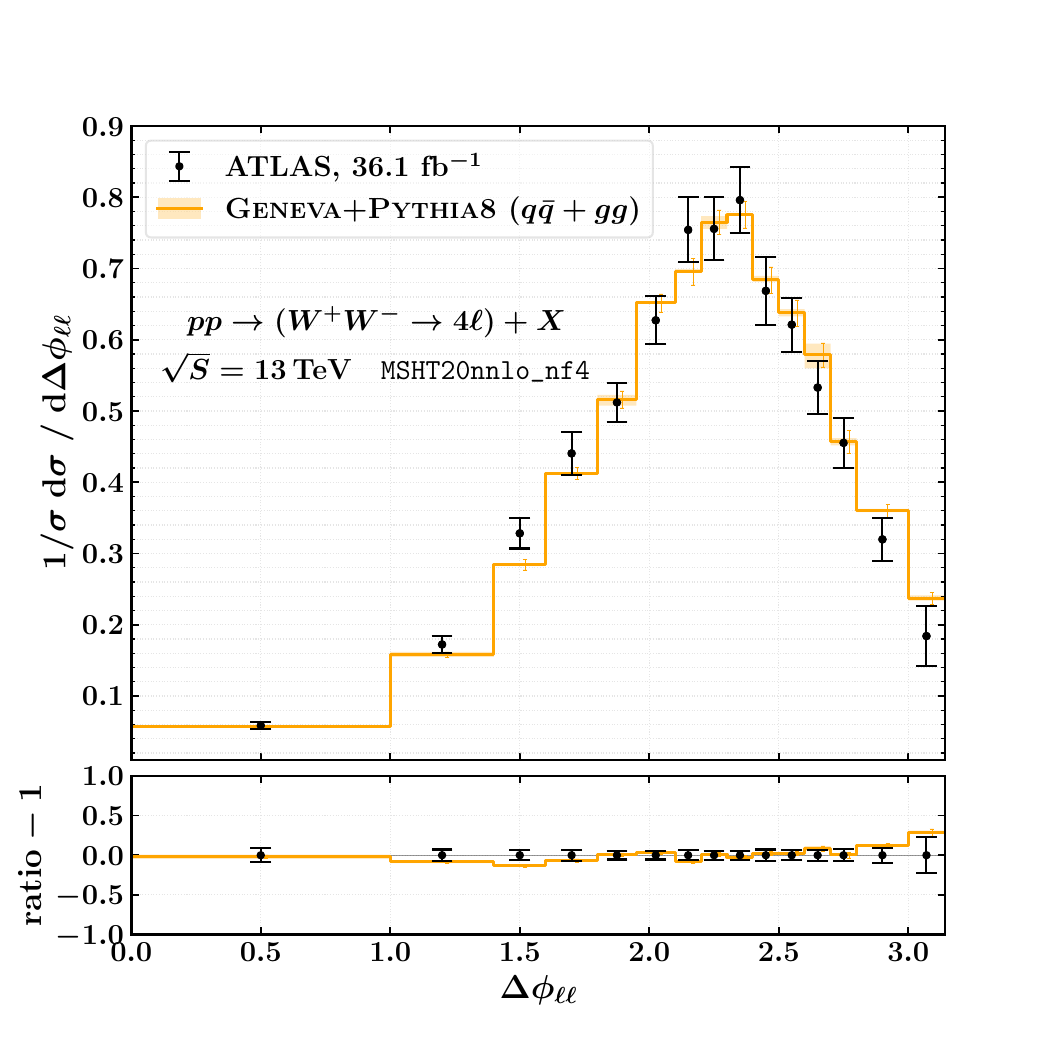} \\
\end{tabular}
\end{center}
\spaceabovefigurecaption
\caption{Comparison of \geneva predictions against ATLAS data for different kinematic distributions. We show the maximum lepton transverse momentum and charged dilepton pair transverse momentum in the top left and right, and the invariant mass and azimuthal separation of the charged lepton pair in the bottom left and right.
\label{fig:ATLASkinematic1}
}
\spacebelowfigurecaption
\end{figure}

\begin{figure}[ht!]
   \centering
   \includegraphics[width=\rescaletwoplots]{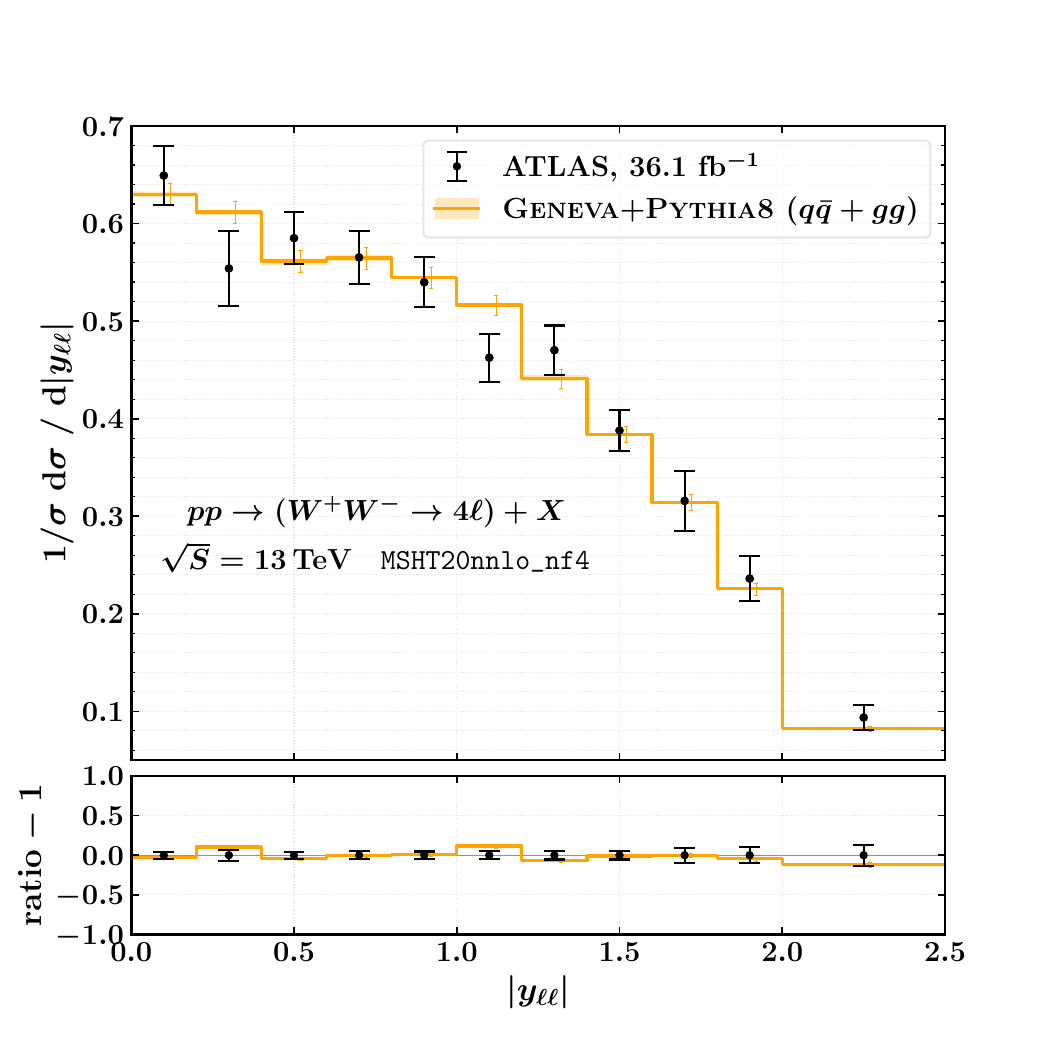}
   \includegraphics[width=\rescaletwoplots]{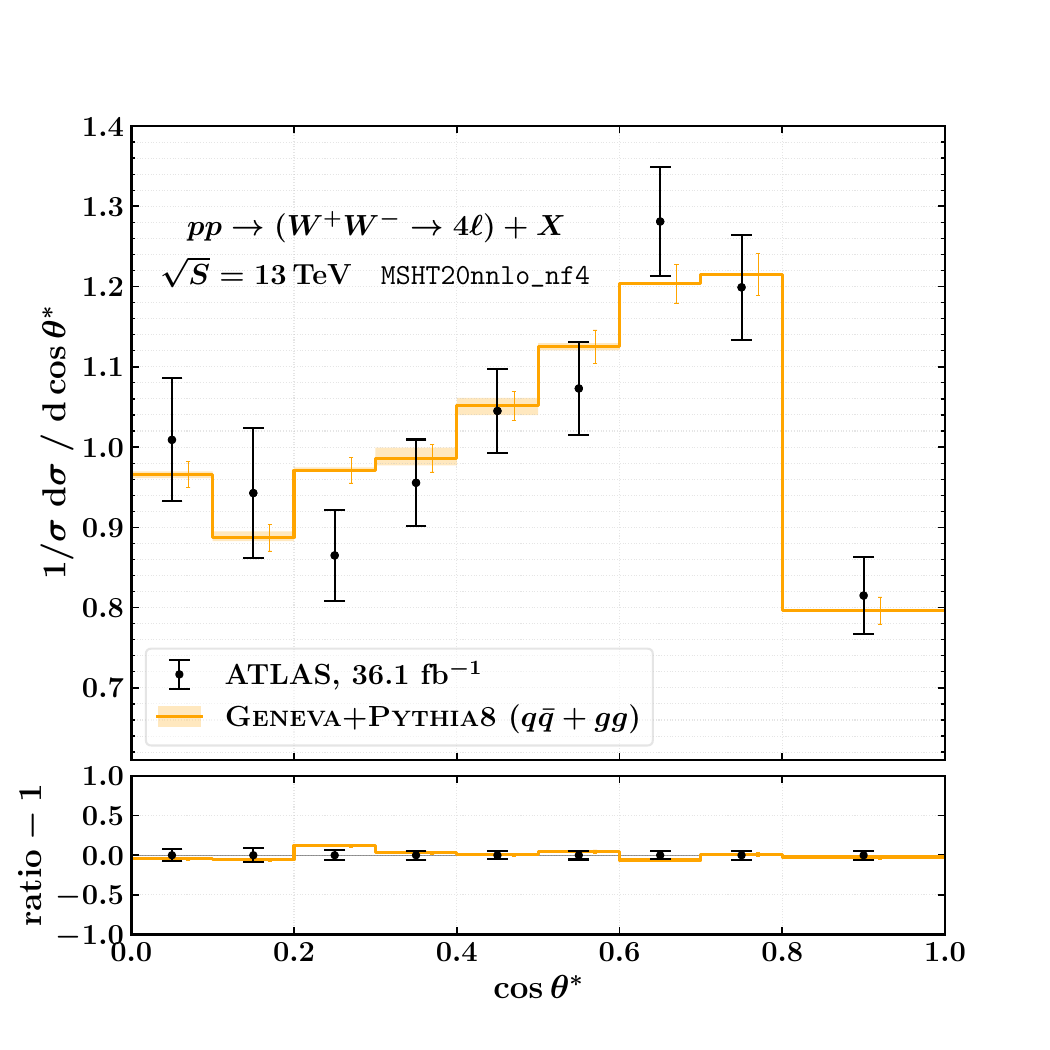}
   \caption{Comparison of \geneva predictions against ATLAS data for different kinematic distributions. We show the rapidity of the charged lepton pair on the left and the cosine of the polar angle in the Collins-Soper frame on the right.
   \label{fig:ATLASkinematic2}
   }
\end{figure}

In \figs{ATLASkinematic1}{ATLASkinematic2} we instead examine comparisons for six kinematic distributions from ATLAS: the transverse momentum of the hardest charged lepton, the transverse momentum, invariant mass, azimuthal separation and rapidity of the charged lepton pair and the cosine of the polar angle between charged leptons in the Collins-Soper frame, defined as
\begin{equation}
\cos \theta^{*} = |\tanh(\Delta \eta_{\ell\ell}/2)|\,.
\end{equation}
 We observe a similar quality of agreement to the CMS case, with all
 distributions being well described by the \geneva predictions excepting the high $p_T$ and $M_{\ell\ell}$ regions of phase space.

\section{Conclusions}
\label{sec:conclusions}
In this work, we have constructed a new `flavour' of the \geneva method, which uses the resummation of jet veto logarithms (rather than the colour-singlet transverse momentum or the $N$-jettiness) to achieve NNLO+PS matching. This is the first NNLO+PS accurate generator to exploit this resolution variable. We have studied the process $pp\to W^+W^-\to 4\ell$, for which jet veto resummation plays an important r\^{o}le due to the large background from $t\bar{t}$ production. Having validated the NNLO (NNLL) accuracy of our fixed-order (resummed) results by comparison with \Matrix and MCFM, we have compared our showered events to data collected by the ATLAS and CMS experiments. We have found good agreement in both cases, with the CMS jet-vetoed cross section being slightly better described by our predictions than the corresponding ATLAS measurement. 

Our predictions for the exclusive $0-$jet cross section currently suffer from rather large theoretical uncertainties. These stem from the inclusion of the $gg$-channel, which is only included at NLL accuracy (since it becomes available at $\mathcal{O}(\alpha_s^2)$ relative to the $q\bar{q}$ channel). In future work, it would be important to increase the theoretical precision by including higher order corrections to this channel, c.f. refs.~\cite{Caola:2015rqy,Grazzini:2020stb, Alioli:2021wpn}. The inclusion of electroweak corrections would also be necessary to improve the description of the high invariant mass/transverse momentum regions of phase space.

Regarding the performance of the code, a major bottleneck is the evaluation of the two-loop hard function. This could in principle be accelerated using a grid-based approach, such as that followed in ref.~\cite{Lombardi:2021rvg}. A similar approach is to use neural networks to learn the functional form of the two-loop expressions and to implement these. We hope to report on developments in this direction in future work. 

In principle, it is also possible to study other colour-singlet production processes with our code. The relatively small size of the hard scale means that jet veto resummation is not particularly important for Drell-Yan ($\sim 90~\GeV$), but for Higgs production in gluon fusion it would certainly be of interest. In addition, processes closely related to colour-singlet production such as Higgs production via vector boson fusion could also be treated with relatively minor modifications to our method.

It is now possible to choose between three versions of \geneva, each using a different resolution variable. It would be interesting to compare the differences between predictions for exclusive observables in a given process, using resummation in either $q_T$, $\mathcal{T}_0$ or $p_T^J$. It would be possible to determine, for example, whether or not using $\textsc{Geneva}_{q_T}$ /$\textsc{Geneva}_{\mathcal{T}}$ results in an equally good description of the exclusive $0-$jet cross section as one obtains using $\textsc{Geneva}_{p_T^J}$. In addition, for processes in which the effects of fiducial power corrections can play a significant r\^{o}le, one might expect large differences due to the sensitivity to the resummation -- varying the resolution variable would then allow an assessment of the reliability of theoretical uncertainty estimates. We leave this to a future study.


\acknowledgments
We thank Andrea Banfi, Alessandro Broggio, Keith Ellis and Rikkert
Frederix for useful discussions, Daniel Gillies for assistance with
running MCFM10 and Stefan Kallweit for providing us with the \Matrix
data used in our comparison. We also thank Pedro Cal, Darren Scott and
Wouter Waalewijn for related collaboration. We are grateful to our
other \scetlib and \geneva collaborators for their ongoing work on the
code, and to the Lawrence Berkeley National Laboratory for hospitality
while part of this work was completed. This project has received
funding from the European Research Council (ERC) under the European
Union's Horizon 2020 research and innovation programme (Grant
agreements 101002090 COLORFREE and 714788 REINVENT). MAL is supported
by the UKRI guarantee scheme for the Marie Sk\l{}odowska-Curie
postdoctoral fellowship, grant ref. EP/X021416/1. We acknowledge the
use of computing resources made available by the Cambridge Service for
Data Driven Discovery (CSD3), part of which is operated by the
University of Cambridge Research Computing on behalf of the STFC DiRAC
HPC Facility (www.dirac.ac.uk). The DiRAC component of CSD3 was funded
by BEIS capital funding via STFC capital grants ST/P002307/1 and
ST/R002452/1 and STFC operations grant ST/R00689X/1. DiRAC is part of
the National e-Infrastructure.

\appendix

\section{The $p_T^J$-preserving mapping}
\label{app:mapping}

In order to perform our NLO$_1$ calculation, we are required to define a mapping from the phase space $\mathrm{d}\Phi_2$ with two final-state partons to the phase space $\mathrm{d}\Phi_1$ with only one. In this work, we wish to enforce that this mapping preserves the transverse momentum of the hardest jet. We begin by dividing $\mathrm{d}\Phi_2$ into two regions which we call
\begin{enumerate}
  \item \textit{ISR}: The two partons belong to different jets.
  \item \textit{FSR}: The two partons belong to the same jet.
\end{enumerate}
We consider a generic configuration $\Phi_2$ with two final-state
partons with transverse momenta $p_{1,T}$ and $p_{2,T}$, rapidities
$y_1$ and $y_2$, and azimuthal angles $\phi_1$ and $\phi_2$
respectively. To establish to which region such a configuration belongs, we need to compare the distance $d_{12}$ between the two
final-state partons to the distances $d_1$ and $d_2$ between each
parton and the beam. Following ref.~\cite{Cacciari:2008gp}, for a jet
clustering algorithm with jet radius $R$, such distances can be
written as
\begin{equation}
  d_{12} = \min\<\L(p_{1,T}^{2p}, p_{2,T}^{2p}\R)
  \frac{\L(y_1-y_2\R)^2 + \L(\phi_1-\phi_2\R)^2}{R^2},
\end{equation}
and
\begin{equation}
  d_1 = p_{1,T}^{2p}
  \qquad\qquad
  d_2 = p_{2,T}^{2p},
\end{equation}
where $p$ is a number that parameterises different possible algorithms:
in particular the choices $p = -1, 0, 1$ correspond to the so-called
$k_T$, Cambridge/Aachen and anti-$k_T$ algorithms. The two partons
then belong to the same jet if $d_{12} < \min\<\L(d_1, d_2\R)$, which
reduces to the condition
\begin{equation}
  \label{ISR_condition}
  \L(y_1-y_2\R)^2 + \L(\phi_1-\phi_2\R)^2 < R^2.
\end{equation}
It is worth noting that the above condition is independent of the
value of $p$, implying that the results discussed in this section
hold for all of the three most popular jet definitions.

Since the phase space $\mathrm{d}\Phi_1$ is parameterised by five independent variables (if we consider the colour singlet to be a single massive particle), we need five conditions to uniquely determine the mapping. In our case, we require that it preserves the following quantities:
\begin{enumerate}
\item Mass of the colour singlet.
\item Rapidity of the colour singlet.
\item Transverse momentum of the hardest jet.
\item Rapidity of the hardest jet.
\item Azimuthal angle of the hardest jet.
\end{enumerate}
Once the momentum of the colour singlet as a whole in the $\Phi_1$
configuration has been determined, the momenta of the particles that
compose it are obtained with a Lorentz transformation of the
corresponding $\Phi_2$ momenta. We note that, in the \textit{ISR}
region where the jet is made by a single parton and thus is massless,
the last three conditions are equivalent to requiring that the mapping
preserves the four-momentum of the hardest parton. In the \textit{FSR}
region instead, since there is only one jet, preserving its
transverse momentum is equivalent to preserving the full four momentum of the colour singlet, which is guaranteed by the conditions above.

\subsection{Integration limits on the splitting variables}

We denote the fractions of the hadronic momenta $P_a$
and $P_b$ carried by the initial-state partons by $x_a$ and $x_b$, the
momenta of the two final-state partons by $p_1$ and $p_2$, and the 
momentum of the colour singlet by $q$. The phase space 
$\mathrm{d}\Phi_2$ can then be written as
\begin{eqnarray}
  \label{pTjet_mapping_dPhi2_starting_point}
  \mathrm{d}\Phi_2 \EI &=& \EI \mathrm{d}x_a \> \mathrm{d}x_b \> \mathrm{d}^4q \>
  \frac{\mathrm{d}^4p_1}{\L(2\pi\R)^3} \> \delta\<\L(p_1^2\R)
  \theta\<\L(p_1^0\R) \frac{\mathrm{d}^4p_2}{\L(2\pi\R)^3} \>
  \delta\<\L(p_2^2\R) \theta\<\L(p_2^0\R)
  \nonumber \\
  && \EI {} \times \delta^4\<\L(q+p_1+p_2-x_aP_a-x_bP_b\R) \>
  \mathrm{d}\Phi_\CS\<\L(q^2\R),
\end{eqnarray}
where, denoting the momenta of the $n$
particles of mass $m_i$ that compose the colour singlet
by $q_i$ for $i = 1, ..., n$,
\begin{equation}
  \mathrm{d}\Phi_\CS\<\L(q^2\R) = \prod_{i=1}^n \L[\frac{\mathrm{d}^4q_i}{\L(2\pi\R)^3}
    \> \delta\<\L(q_i^2-m_i^2\R) \theta\<\L(q_i^0\R)\R] \L(2\pi\R)^4
  \delta^4\<\L(\sum_{i=1}^nq_i-q\R).
\end{equation}
To simplify the expression of
eq.~(\ref{pTjet_mapping_dPhi2_starting_point}) we notice that, for a
generic momentum $p$, $\mathrm{d}^4p$ can be written as
\begin{equation}
  \mathrm{d}^4p = \frac{1}{2} \> \mathrm{d}p^2 \> \mathrm{d}y \> \mathrm{d}^2\vec{p}_T = \frac{1}{4} \> \mathrm{d}p^2
  \> \mathrm{d}y \> \mathrm{d}p_T^2 \> \mathrm{d}\phi,
\end{equation}
where $p^2$, $y$, $p_T$ and $\phi$ are respectively the virtuality,
rapidity, transverse momentum and azimuthal angle of $p$, while
$\delta^4\<\L(p\R)$ can be expressed as
\begin{equation}
  \delta^4\<\L(p\R) = 2 \> \delta\<\L(p^-\R) \delta\<\L(p^+\R)
  \delta^2\<\L(\vec{p_T}\R).
\end{equation}
These identities allow us to express $\mathrm{d}\Phi_2$ as
\begin{eqnarray}
  \mathrm{d}\Phi_2 \EI &=& \EI \frac{\mathrm{d}m_\CS^2 \>
    \mathrm{d}y_\CS}{S} \> \frac{\mathrm{d}p_{1,T}^2 \> \mathrm{d}y_1
    \> \mathrm{d}\phi_1}{4\L(2\pi\R)^3} \> \frac{\mathrm{d}p_{2,T}^2
    \> \mathrm{d}y_2 \> \mathrm{d}\phi_2}{4\L(2\pi\R)^3} \>
  \mathrm{d}\Phi_\CS\<\L(m_\CS^2\R) \theta\<\L(\sqrt{S} - e^{y_1}
  p_{1,T} - e^{y_2} p_{2,T} \R.
  \nonumber \\
  && \EI \L. {} - e^{y_\CS} \sqrt{m_\CS^2 + p_{1,T}^2 + p_{2,T}^2 +
    2p_{1,T}p_{2,T}\cos\<\L(\phi_2-\phi_1\R)}\R) \theta\<\L(\sqrt{S} -
  e^{-y_1} p_{1,T} - e^{-y_2} p_{2,T} \R.
  \nonumber \\
  && \EI \L. {} - e^{-y_\CS} \sqrt{m_\CS^2 + p_{1,T}^2 + p_{2,T}^2 +
    2p_{1,T}p_{2,T}\cos\<\L(\phi_2-\phi_1\R)}\R),
  \nonumber \\
\end{eqnarray}
where $m_\CS$ and $y_\CS$ are respectively the mass and rapidity of
the colour singlet and $p_{1,T}$, $y_1$, $\phi_1$, $p_{2,T}$, $y_2$
and $\phi_2$ are the transverse momentum, rapidity and azimuthal angle
for the two partons.

The above expression is particularly useful because it expresses the
differential phase space $\mathrm{d}\Phi_2$ in terms of five variables that are
preserved by the mapping ($m_\CS$, $y_\CS$, $p_{1,T}$, $y_1$ and
$\phi_1$), which we can use to parameterise the projected phase space
$\mathrm{d}\Phi_1$, together with the variable we are resumming
($p_{2,T}$). In order to generate events distributed according to the
resummed $p_{2,T}$ spectrum, we then need to compute the integration
limits of the remaining two variables $y_2$ and $\phi_2$, which are
obtained by imposing that the arguments of the two $\theta$ functions
are positive and correspond to the common solutions of the two sets of
inequalities (parameterised by the two signs)
\begin{equation}
  \L\{
  \begin{array}{l}
    \DS \sqrt{S} - e^{\pm y_1} p_{1,T} - e^{\pm y_2} p_{2,T} > 0\, ,
    \Biggwhite \\
    \DS \cos\<\L(\phi_2-\phi_1\R) < \frac{e^{\mp 2y_\CS} \L(\sqrt{S} -
      e^{\pm y_1} p_{1,T} - e^{\pm y_2} p_{2,T}\R)^2 - m_\CS^2 -
      p_{1,T}^2 - p_{2,T}^2}{2 p_{1,T} p_{2,T}}.
    \Biggwhite
  \end{array}
  \R.
\end{equation}
The second of the above inequalities immediately gives the integration
limits on $\phi_2$, while those on $y_2$ are found by imposing that
said inequality on the variable $\phi_2$ has a non-empty set of
solutions and are given by
\begin{equation}
  y_2 \lessgtr \log\<\L(\frac{\sqrt{S} - e^{\pm y_1} p_{1,T} - e^{\pm
      y_\CS} \sqrt{m_\CS^2 + \L(p_{1,T}-p_{2,T}\R)^2}}{p_{2,T}}\R).
\end{equation}

At this point, we need to keep into account that we want to use this
mapping only in the \textit{ISR} region, which, from
eq.~(\ref{ISR_condition}), imposes the further constraint on $\phi_2$
\begin{equation}
  \L(\phi_2-\phi_1\R)^2 > R^2 - \L(y_2-y_1\R)^2.
\end{equation}
Finally, in order to introduce a variable $z$ that we will use in the
expressions of the splitting functions used to spread the resummed
spectrum over the entire $\mathrm{d}\Phi_2$ phase space, we need to further
divide the \textit{ISR} region into two subregions that we call
\begin{enumerate}
\item \textit{ISRA}: The region where
  \begin{equation}
    y_2 > y_\CS,
  \end{equation}
  where we define
  \begin{equation}
    z = \frac{\sqrt{S} \b{x}_a}{\sqrt{S} \b{x}_a + p_2^-},
  \end{equation}
  (with the barred variables representing quantities in the underlying Born configuration) so that
  \begin{equation}
    y_2 = \log\<\L(\frac{\sqrt{S} \b{x}_a}{p_{2,T}} \frac{1-z}{z}\R).
  \end{equation}
\item \textit{ISRB}: The region where
  \begin{equation}
    y_2 < y_\CS,
  \end{equation}
  where we define
  \begin{equation}
    z = \frac{\sqrt{S} \b{x}_b}{\sqrt{S} \b{x}_b + p_2^+},
  \end{equation}
  so that
  \begin{equation}
    y_2 = - \log\<\L(\frac{\sqrt{S} \b{x}_b}{p_{2,T}}
    \frac{1-z}{z}\R).
  \end{equation}
\end{enumerate}
The phase space $\mathrm{d}\Phi_2$ expressed in terms of $z$ then reads
\begin{eqnarray}
  \mathrm{d}\Phi_2 \EI &=& \EI \frac{\mathrm{d}m_\CS^2 \> \mathrm{d}y_\CS}{S} \>
  \frac{\mathrm{d}p_{1,T}^2 \> \mathrm{d}y_1 \> \mathrm{d}\phi_1}{4\L(2\pi\R)^3} \>
  \frac{\mathrm{d}p_{2,T}^2 \> \mathrm{d}z \> \mathrm{d}\phi}{4\L(2\pi\R)^3 z\L(1-z\R)} \>
  \mathrm{d}\Phi_\CS\<\L(m_\CS^2\R),
\end{eqnarray}
where we dropped the $\theta$ functions for ease of notation and
renamed $\phi_2$ to $\phi$ to make contact with the notation from
previous works.

\section{The splitting function kernels}
\label{app:splitting_kernels}

For a complete discussion of the implementation of the splitting
functions $\mathcal{P}_{N \to N+1}\<\L(\Phi_{N+1}\R)$ in \Geneva{} we
refer the reader to section 3.1 of ref.~\cite{Alioli:2023har}. In this
appendix, we will limit ourselves to introducing the minimal notation
needed to present the new kernels used in this paper to improve the
description of the fully differential resummed contribution when the
transverse momentum is used as resolution variable. Following the
prescription of ref.~\cite{Alioli:2023har}, in order to fulfil the
condition presented in eq.~\ref{eq:cPnorm}, we choose the splitting
functions $\mathcal{P}_{N \to N+1}\<\L(\Phi_{N+1}\R)$ such that they
depend on the mother and sister indices of the QCD splitting and
vanish in the unprojectable $\Phi_{N+1}$ configurations
\begin{equation}
\mathcal{P}_{N \to N+1}\<\L(\Phi_{N+1}\R) = \Bigg\{
  \begin{array}{l l}
  0 & \mbox{if $\Phi_{N+1}$ is unprojectable}
  \\
  \mathcal{P}_{kj} (\Phi_N, \Tau_N, z, \phi) \quad & \mbox{if $\Phi_N
    \to \Phi_{N+1}$ via the $k \to i + j$ splitting,}
  \end{array}
\end{equation}
where we define
\begin{eqnarray}
  && \mathcal{P}_{kj} \<\L(\Phi_N, \Tau_N, z \R) =
  \\
  && \quad \frac{\DS f_{kj}\<\L(\Phi_N, \Tau_N, z\R)}{\DS
    \sum_{k'=1}^{N+2} \int_{z^\MIN_{k'}\<\L(\Phi_N,
      \Tau_N\R)}^{z^\MAX_{k'}\<\L(\Phi_N, \Tau_N\R)} \df z' \>
    J_{k'}\<\L(\Phi_N, \Tau_N, z'\R) \Delta \phi_{k'}\<\L(\Phi_N, \Tau_N,
    z'\R) \sum_{j'=1}^{n^{\rm split}_{k'}} f_{k'j'}\<\L(\Phi_N,
    \Tau_N, z'\R)},
  \nonumber
\end{eqnarray}
where
\begin{equation}
  J_k\<\L(\Phi_N, \Tau_N, z\R) = \L.\frac{\df\Phi_{N+1}}{\df\Phi_N \>
    \df\Tau_N \> \df z \> \df\phi}\R|_k,
\end{equation}
$|_k$ indicates a fixed value of $k$ and $\Delta \phi_k\<\L(\Phi_N,\Tau_N,z\R) =
\phi^\MAX_k\<\L(\Phi_N,\Tau_N,z\R) -
\phi^\MIN_k\<\L(\Phi_N,\Tau_N,z\R)$.
To motivate the choice of the splitting function kernels we use, we
follow the reasoning of ref.~\cite{Alioli:2023har} and consider the $k
\to i + j$ splitting connecting the Born matrix element
$\mathcal{B}_0$ and the real matrix element $\mathcal{B}_1$ in the
case of colour singlet production in hadron-hadron collisions. We
introduce the FKS variables $\xi = 2 \> E / \sqrt{s}$ and $y = \cos
\theta$, where $s$ is the squared partonic centre-of-mass energy and
$E$ and $\theta$ are the energy of the emitted parton and the angle
between the emitted and the right-moving incoming parton in the
partonic centre-of-mass frame. In the soft limit of the emitted
particle $i$, we have
\begin{equation}
  \label{eq:0to1_soft_limit}
  \lim_{\xi \to 0} \mathcal{B}_1 =
  \frac{64\pi\alphaS\<\L(\mu_R\R)}{Q^2} \frac{C_{k}}{\xi^2\L(1-y^2\R)}
  \> \mathcal{B}_0,
\end{equation}
where $C_{k} = \CF$ for the quark-initiated processes and $C_{k} =\CA$
for the gluon-initiated, while in the azimuthally averaged collinear
limit between particles $i$ and $j$, we have
\begin{equation}
  \label{eq:0to1_collinear_limit}
  \lim_{y \to \pm 1} \mathcal{B}_1 =
  \frac{16\pi\alphaS\<\L(\mu_R\R)}{Q^2} \frac{1-\xi}{\xi\L(1\mp y\R)}
  \> \h{P}_{jk}\<\L(1-\xi\R) \mathcal{B}_0,
\end{equation}
where $y \to 1$ and $y \to -1$ represent the collinear limits with
respect to incoming parton $a$ and $b$ respectively, and
$\h{P}_{jk}\<\L(1-\xi\R)$ are the Altarelli-Parisi splitting
functions. If the colour singlet production process is quark-initiated
or has only scalar particles in the final state, the above expressions
also hold prior to averaging over the azimuthal angle. At this point
we deviate from the discussion in ref.~\cite{Alioli:2023har} and write
$p_T^2$ and $z$ in terms of $\xi$ and $y$, obtaining
\begin{eqnarray}
    \DS p_T^2 \EI &=& \EI \frac{Q^2}{4} \> \frac{\xi^2}{1-\xi}
    \L(1-y^2\R)
    \nonumber \\
    \DS z \EI &=& \EI \L(1 + \frac{\xi\L(1\pm y\R)}{2\sqrt{1-\xi}}
    \sqrt{\frac{2-\xi\L(1\mp y\R)}{2-\xi\L(1\pm y\R)}}\R)^{-1}.
\end{eqnarray}
In the soft limit, the above expressions simplify to
\begin{eqnarray}
  p_T^2 \EI &\to& \EI \frac{Q^2}{4} \> \xi^2 \L(1-y^2\R)
  \\
  z \EI &\to& \EI 1 - \frac{\xi}{2} \L(1 \pm y\R),
\end{eqnarray}
while in the collinear limit they become
\begin{eqnarray}
  p_T^2 \EI &\to& \EI \frac{Q^2}{2} \> \frac{\xi^2}{\L(1-\xi\R)}
  \L(1 \mp y\R)
  \\
  z \EI &\to& \EI 1-\xi.
\end{eqnarray}
In order to reproduce the correct soft and collinear limits of
eqs.~(\ref{eq:0to1_soft_limit}) and~(\ref{eq:0to1_collinear_limit})
in this case, the splitting kernels presented in eq. (3.9) of
ref.~\cite{Alioli:2023har} are substituted by
\begin{equation}
  f_{kj}\<\L(\Phi_N, p_T^2, z\R) =
  \frac{8\pi\alphaS\<\L(\mu_R\R)}{p_T^2} f_a^A\<\L(x_a,\mu_F\R)
  f_{b}^B\<\L(x_b,\mu_F\R) \L(1-z\R) \h{P}_{jk}\<\L(z\R),
\end{equation}
where $a$ and $b$ are the initial-state partons, $\alphaS\<\L(\mu_R\R)$
is the strong coupling evaluated at the renormalisation scale $\mu_R$,
and $f_i^H\<\L(x_i,\mu_F\R)$ is the PDF of the parton $i$ in the hadron
$H$ evaluated at longitudinal momentum fraction $x_i$ and
factorisation scale $\mu_F$.


\addcontentsline{toc}{section}{References}
\bibliographystyle{jhep}
\bibliography{genevaptveto}

\end{document}